\documentclass[journal]{IEEEtran}
\usepackage[noadjust]{cite}

\ifCLASSINFOpdf
\else
\fi

\usepackage{amsmath}
\usepackage{amssymb}
\usepackage{hyperref}
\usepackage{subcaption}
\usepackage{graphicx}
\usepackage{cleveref}
\Crefname{figure}{Fig.}{Figs.}
\usepackage{bm}
\usepackage{tensor}
\usepackage{my-Equations}       
\usepackage{amsthm}
\usepackage{tikz}
\usepackage[normalem]{ulem}
\usepackage[shortlabels]{enumitem}

\usepackage{xr-hyper}
\makeatletter
\newcommand*{\addFileDependency}[1]{
  \typeout{(#1)}
  \@addtofilelist{#1}
  \IfFileExists{#1}{}{\typeout{No file #1.}}
}
\makeatother

\newcommand*{\myexternaldocument}[1]{%
    \externaldocument{#1}%
    \addFileDependency{#1.tex}%
    \addFileDependency{#1.aux}%
}
\myexternaldocument{supplementary_material}

\theoremstyle{definition}
\newtheorem{theorem}{Theorem}[]

\theoremstyle{definition}
\newtheorem{definition}[theorem]{Definition}

\theoremstyle{definition}
\newtheorem{proposition}[theorem]{Proposition}

\theoremstyle{definition}
\newtheorem{lemma}[theorem]{Lemma}

\theoremstyle{definition}

\theoremstyle{remark}
\newtheorem{remark}{Remark}

\usepackage{pdfpages} 
\usepackage{pgffor} 

\makeatletter
\AtBeginDocument{\let\LS@rot\@undefined}
\makeatother

\def\supplementfilename{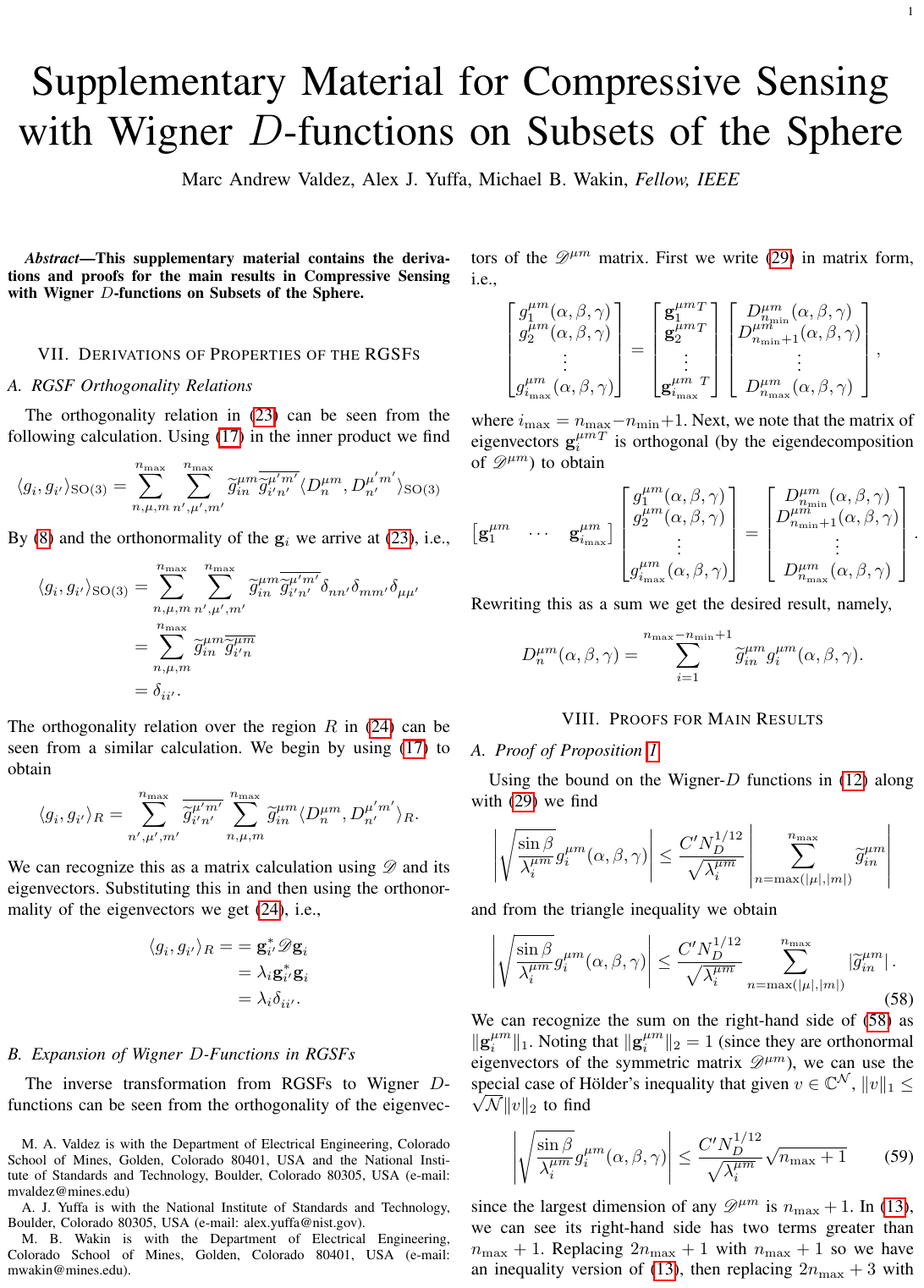}

\pdfximage{\supplementfilename}
\def\numbersupplementpages{\the\pdflastximagepages}

\newif\ifarXiv
\arXivtrue 
\begin{document}
%
\title{Compressive Sensing with Wigner $D$-functions on Subsets of the Sphere}
%
%
%

\author{Marc~Andrew~Valdez,
        Alex~J.~Yuffa,
        Michael~B.~Wakin,~\IEEEmembership{Fellow,~IEEE}%
\thanks{M.~A.~Valdez is with the Department of Electrical Engineering, Colorado School of Mines, Golden, Colorado 80401, USA and the National Institute of Standards and Technology, Boulder, Colorado 80305, USA (e-mail: mvaldez@mines.edu)}%
\thanks{A.~J.~Yuffa is with the National Institute of Standards and Technology, Boulder, Colorado 80305, USA (e-mail: alex.yuffa@nist.gov).}%
\thanks{ M.~B.~Wakin is with the Department of Electrical Engineering, Colorado School of Mines, Golden, Colorado 80401, USA (e-mail: mwakin@mines.edu).}%
}        
\maketitle

\begin{abstract}
In this paper, we prove a compressive sensing guarantee for restricted measurement domains on the rotation group, $\sot$. We do so by first defining Slepian functions on a measurement sub-domain $R$ of the rotation group $\sot$. Then, we transform the inverse problem from the measurement basis, the bounded orthonormal system of band-limited Wigner $D$-functions on $\sot$, to the Slepian functions in a way that limits increases to signal sparsity. Contrasting methods using Wigner $D$-functions that require measurements on all of $\sot$, we show that the orthogonality structure of the Slepian functions only requires measurements on the sub-domain $R$, which is select-able. Due to the particulars of this approach and the inherent presence of Slepian functions with low concentrations on $R$, our approach gives the highest accuracy when the signal under study is well concentrated on $R$. We provide numerical examples of our method in comparison with other classical and compressive sensing approaches. In terms of reconstruction quality, we find that our method outperforms the other compressive sensing approaches we test and is at least as good as classical approaches but with a significant reduction in the number of measurements.
\end{abstract}


%
\IEEEpeerreviewmaketitle

\section{Introduction}
%
%
%
%


\IEEEPARstart{I}{n} antenna design and metrology, characterization of an antenna's (or antenna array's) far-field radiation profile is of the utmost importance. This is especially true as 5G and beyond devices are developed with increasing power, controllability, and potential for secondary lobes to damage other devices when in high power regimes \cite{rappaport_wireless_2019}. While numerical simulations give an ideal view of a device under test (DUT) far-field profile, it is often necessary to physically characterize this far-field profile to ensure manufactured devices actually meet design requirements. The now-canonical approach to characterizing a DUT far-field profile is termed spherical near-field to far-field (NF2FF) transformation~\cite{hansen_spherical_1988}, which allows near-field measurements to be transformed to a far-field characterization. Unfortunately, measurements for characterizing DUTs are very time-consuming and require large amounts of data, especially when higher frequencies like those used in 5G technology are involved~\cite{hansen_spherical_1988, loschenbrand_fast_2016, fuchs_fast_2017}. Adding to the challenge, physical limitations of the measurement apparatus can prohibit measurements from being taken in certain regions of space. For example, when a DUT is placed on a support structure with its main beam pointing up, it may not be possible to accurately measure the bottom portion of the near field.

Spherical NF2FF characterizations use near-field measurements on a sphere enclosing the DUT to determine its field expansion coefficients in the band-limited spherical wave-function (SW) basis. When characterizing electromagnetic (EM) devices, vector SWs are used, and when characterizing an acoustic device like a loudspeaker, scalar SWs are used (in-air acoustics require only p-waves). The coefficients in these bases are sometimes called spherical mode coefficients. With these coefficients in hand, it is possible to determine the far-field radiation profile. A common and accurate sampling approach for spherical NF2FF transformations is the equiangular sampling pattern with fixed polar and azimuthal step sizes~\cite{hansen_spherical_1988, wittmann_probe-corrected_1992}. This approach uses classic Nyquist theory. In the best case, using a probe that is only sensitive to the lowest SW modes (a $\mu=\pm1$ probe for EM applications, or an axisymmetric $\mu=0$ probe in acoustics), the Nyquist-based approach requires a number of measurements that is quadratic in the band-limit~\cite{hansen_spherical_1988}. Even more measurements are needed for high accuracy in the presence of noise. Interestingly, this method can cope with restricted measurement regions~\cite{hansen_spherical_1988, yaccarino_application_1991}. In particular, the Nyquist-based method copes with measurement restrictions by using measurements from available positions and padding the remainder of the sphere with zeros. The cost of this zero padding is that the field reconstruction is accurate only in a subset of the measured region~\cite{hansen_spherical_1988, yaccarino_application_1991}. However, this approach can still require many measurements and the need to pad with zeros in unmeasurable regions is highly unfavorable~\cite{wittmann_using_2012, saccardi_truncation_2017}. 

Compressive Sensing (CS) has proved to be a powerful tool for reducing the number of measurements needed to acquire signals, provided they are sparse~\cite{candes_decoding_2005, candes_robust_2006, candes_near-optimal_2006, rauhut_random_2007, rudelson_sparse_2008,  rauhut_sparse_2012, foucart_mathematical_2013}. Sparse signals are those that, in some representation, contain a small number of nonzero values. It is well-known that the SW coefficients tend to be sparse and concentrated near the bottom end of the band when the coordinate system for the SWs is centered on the phase center of the DUT~\cite{loschenbrand_fast_2016, culotta-lopez_compressed_2018}. Thus, when there are no measurement restrictions, CS-based random sampling offers a way to reduce the number of measurements needed for accurate field reconstructions~\cite{loschenbrand_fast_2016,fuchs_fast_2017, culotta-lopez_compressed_2018, cornelius_compressed_2016, fuchs_compressive_2018, bangun_sensing_2020}. However, none of the recent work in this vein has developed an approach that accounts for measurement restrictions. Motivated by CS in NF2FF characterizations, in this paper, we develop a general theory for CS that allows for measurements to be taken on a restricted domain and can give field reconstructions within that measured region\footnote{\Cref{sec:background,sec:cs_stuff,sec:main_res} provide the general derivations and statements of results. Due to their technical nature, proofs are provided as supplementary material.}.

Every CS recovery problem involves a measurement matrix relating the observed data to the unknown vector of interest. When this matrix satisfies certain conditions, one can ensure that CS recovery algorithms such as quadratically constrained basis pursuit (QCBP), a form of $\ell_1$ minimization, will achieve accurate and robust recovery. Satisfying the Restricted Isometry Property (RIP) \cite{foucart_mathematical_2013} is one well-known sufficient condition for a measurement matrix. An important class of measurement matrices that satisfy the RIP are matrices generated by random sampling in a Bounded Orthonormal System (BOS). Letting $\{f_j(x)\}$ denote a set of uniformly bounded orthonormal functions, taking random samples at locations $\{x_i\}$ yields a measurement matrix with entries $\Phi_{ij} = f_j(x_i)$. In such a setting, the number of measurements to guarantee accurate CS recovery scales with the square of the uniform bound. Much work has been devoted to showing that certain classes of special functions can be used as BOSs~\cite{rauhut_sparse_2012, bangun_sensing_2020}. The most important of these special functions for NF2FF characterizations is the band-limited Wigner $D$-functions \cite{bangun_sensing_2020}.

When collecting measurements for NF2FF characterizations, the measurements must be corrected for the probe's response to the incident field (probe correction). Thus, these measurements are taken in the probe's coordinate system. As a result, the measurements can be represented as a series of Wigner $D$-functions whose coefficients are a combination of the transformation constants and the SW coefficients \cite{hansen_spherical_1988}; see Section~\ref{sec:nf2ff}. The Wigner $D$-functions are an irreducible representation of the rotation group $\mrm{SO}(3)$ \cite{varshalovich_quantum_1988}; their arguments represent the position on the sphere and the polarization angle for a measurement. In \cite{bangun_sensing_2020}, Bangun \tit{et al}. showed that with a sufficient number of measurements, the measurement matrix created from the BOS of band-limited Wigner $D$-functions satisfies the RIP with high probability. That work put CS for NF2FF on a solid theoretical footing. However, since the domain for the Wigner $D$-functions is all of $\mrm{SO}(3)$, the theoretical guarantee in~\cite{bangun_sensing_2020} holds only when one is able to collect random measurements from the entire domain. For cases involving device support structures or other exclusions, there is as of yet no guarantee that CS can be applied to measurements restricted to a portion of a sphere.

\subsection{Approach and Contributions}
In this work, we avoid the limitations present when using the Wigner $D$-functions by transforming the measurement basis to a different set of bounded orthonormal functions on the rotation group, namely, a certain set of Slepian functions. Slepian functions were originally studied as the solution to the spectral concentration problem~\cite{slepian_prolate_1961, landau_prolate_1961, landau_prolate_1962, simons_scalar_2015}, i.e., the problem of finding the functions within a certain band-limit that are maximally localized to a spatial (or temporal) region, $R$. More recently, Slepian functions have been used in various areas of signal processing, from novel wavelet constructions \cite{roddy_slepian_2021} to new tools in signal representation on the sphere \cite{aslam_linear_2021, nafees_differential_2020}. Slepian functions form an orthogonal basis for band-limited $L_2$ functions on their full domain $\mcl{D}$ as well as on $R \subset \mcl{D}$ \cite{simons_scalar_2015}. In this paper, we show this latter property enables CS recovery guarantees from measurements on a restricted domain $R$. However, this benefit comes at a cost. Ensuring the unit normalization of the Slepian functions on $R$ increases their uniform bound, particularly due to ``trailing'' Slepian functions which are poorly localized to $R$. This problem can be mitigated, though, if the signals one wishes to recover using CS are reasonably localized to the measurable domain $R$. 

Mathematically, the approach we take to theoretically guarantee CS recovery on the restricted domain $R$ is as follows. We begin with the full $\mrm{SO}(3)$ CS problem,
\begin{equation}\label{eq:initial_cs_linear_problem}
    w = \Phi a + \eta,
\end{equation}
where $w$ is the vector of measurements of a function $w(\alpha, \beta, \gamma)$ on $\mrm{SO}(3)$, $\Phi\in \mbb{C}^{M\times N}$ with $M<N$ is the measurement matrix of the band-limited Wigner $D$-functions sampled at the corresponding locations $(\alpha, \beta, \gamma)$, and $a$ is the vector of coefficients for the band-limited Wigner $D$-function series. The direct CS approach to this problem, as done in ~\cite{bangun_sensing_2020}, requires collecting enough random measurements from $\mrm{SO}(3)$ and then using QCBP ($\ell_1$ minimization) to solve for the coefficient vector $a$. The need for random measurements from $\mrm{SO}(3)$ derives from the fact that the Wigner $D$-functions are a bounded orthonormal basis for $\mrm{SO}(3)$ and CS theory for BOSs requires measurements from the full domain of orthogonality~\cite{bangun_sensing_2020}. As such, the problem in \eqref{eq:initial_cs_linear_problem} is not viable for CS when the measurements are limited to $R \subset \mcl{D}$.

To avoid this orthogonality problem, we derive an invertible transformation matrix, $V$, that maps $\Phi$ into a new measurement matrix $\Phi'$ containing Slepian functions on $\mrm{SO}(3)$:
\begin{equation}
    \label{eq:Phi_a_remap}
    w = \Phi V^{-1} V a +\eta= \Phi' a' + \eta.
\end{equation}
The Slepian basis in $\Phi' = \Phi V^{-1}$ is orthogonal on $R$ as well as on $\mrm{SO}(3)$ and $a' = V a$. Moreover, if $R$ is a latitudinal belt on $\mrm{SO}(3)$ where only the polar angle of the measurements is restricted, the Slepian functions are bounded similar to the Wigner $D$-functions but with a factor $\lambda \in (0,1)$ of the Slepian concentration in the denominator. Importantly, due to symmetry in the azimuthal and polarization angles of measurement, we show that when $V$ is applied to $a$, it has a bounded effect on increasing the sparsity level of the problem.

Due to the factor $\lambda$ of the concentration, the bound on the Slepian basis in $\Phi'$ can become too large for CS guarantees to be practical because there are Slepian functions whose concentration is near zero. To avoid this, we partition the basis into two sets: one set, $\Phi_1'$, containing the Slepian functions with large concentrations on $R$ and the other set, $\Phi_2'$, containing Slepian functions with small concentrations on $R$.  This partitioning re-expresses \eqref{eq:Phi_a_remap} as
\begin{equation}
    w = \Phi_1' a_1' +\Phi_2' a_2'  + \eta.
\end{equation}
In cases where $w(\alpha, \beta, \gamma)$ has little energy in the complement of $R$, $R^{\mrm{c}}$, we can bound the contribution of $\Phi_2'a_2'$ and group it with the noise term $\eta$. This results in the formulation
\begin{equation}
    w = \Phi_1' a_1' + \eta',
\end{equation}
where $a_1'$ are the coefficients of the Slepian functions in $\Phi_1'$ and $\eta' = \eta + \Phi_2' a_2'$ is a noise term bounded by the sum of the noise $\eta$ and the energy of the $w(\alpha, \beta, \gamma)$ in $R^{\mrm{c}}$, the unmeasurable part of $\mrm{SO}(3)$. Thanks to the uniform bound and orthonormality on $R$ of the Slepian functions in $\Phi_1'$, this new CS problem can be solved using standard guarantees for BOSs with measurements only taken on $R$. We emphasize that this partitioning of the problem implies that solutions to this inverse problem will, like classical zero padding, only estimate field values well within $R$; values in $R^{\mrm{c}}$ are not assumed to be correct.

In practice, for functions well-localized on the measurable domain $R$, our approach is very competitive against other methods. Numerical experiments show that it is possible to achieve similar CS enabled reductions in measurements as were seen over the full domain in~\cite{bangun_sensing_2020}. In particular, our example shows that using only around $60\%$ as many measurements as the classical restricted FFT method results in similar or even better reconstruction performance and does not require an equiangular sampling pattern. Moreover, when compared to various \textit{ad hoc} restricted measurement adaptations of the work in \cite{bangun_sensing_2020}, our method performs better when the measurement restrictions are sizable. In contrast, when measurements can be taken over all of or nearly all of the domain, CS using Wigner $D$-functions as in~\cite{bangun_sensing_2020} is more favorable. This difference in performance is due to the difference in approach when constructing the Slepian function BOS on $\sot$, which prioritizes the control of function orthogonality at the cost of an increased function bound. However, it is this control over orthogonality relations when using Slepian functions that enables theoretical recovery guarantees and practical gains in terms of the required number of measurements in the restricted measurement cases. 


\subsection{Notation}
In this paper, we use the following notation and conventions. The sum $\sum_{n,m,\mu}^{n_{\max}}$ is used to mean $\sum_{n=n_0}^{n_{\max}}\sum_{m=-n}^n \sum_{\mu=-n}^n$, where $n_0=1$ for EM applications and $n_0=0$ for acoustics applications. We use $\mi=\sqrt{-1}$ as the unit imaginary number. An over-line represents complex conjugation, e.g., $\overline{a}$. We represent the Hermitian conjugate of a vector or matrix with a $^*$, e.g., $a^* = \overline{a^T}$, where $T$ denotes the transpose. The norm $\|\cdot\|_p$ with $p\geq 1$ is the standard $\ell_p$ vector norm. $\|a\|_0$ is the $\ell_0$ ``norm'' counting the number of nonzero entries in a vector $a$. $\|\cdot\|_\infty$ is either the $\ell_\infty$ or $L_\infty$ norm, which should be discernible from the context. As usual, for a vector $a$, $\|a\|_\infty = \max_i(|a_i|)$, and for a function $f$, $\|f\|_\infty = \inf\{c\geq0:|f(x)|\leq c\; \text{for almost every} \; x\}$. For Euler rotations, we use the $zyz'$ and passive transformation conventions. $\mcl{U}(a,b)$ is the uniform distribution on the interval $[a,b]$. We use i.i.d.\ to abbreviate independently and identically distributed.

\section{Background}\label{sec:background}

\subsection{Near Field to Far-field Antenna Measurements} \label{sec:nf2ff}
One of the main approaches to characterize a DUT far-field radiation pattern is to use a spherical NF2FF transformation. In the NF2FF transformation, near-field measurements are taken on a sphere enclosing the DUT and used to solve for SW coefficients in a band-limited SW series. In theory, an infinite series expansion for DUT's field should be used. In practice, terms above a band-limit, $n_{\max}$, are so small that they can be ignored. In EM, this band-limit is imposed by the antenna's electrical size and typically set to $n_{\max} = kr_a + 10$, where $k$ is the wavenumber of the emitted radiation and $r_a$ is the radius of the smallest sphere circumscribing the radiating parts of the antenna. A similar procedure is used in acoustics. Once the expansion coefficients up to $n_{\max}$ are determined, the radiated field of the DUT can be computed anywhere outside of the minimum circumscribing sphere.

In EM applications it has become recognized that the SW coefficients are sparse/compressible when an antenna's phase center coincides with the center of the near-field measurement sphere and the antenna's main beam is properly oriented~\cite{loschenbrand_fast_2016, culotta-lopez_compressed_2018}. When the DUT field is measured, however, the SW coefficients are not measured directly. Instead, if multiple scattering is negligible, the probe's response to the incident field is measured. This response is related to the original field SW coefficients by rotation and translation transformations as well as probe correction if a non-ideal probe is used~\cite{hansen_spherical_1988, wittmann_probe-corrected_1992}. With a fixed translation distance, this transformation results in a series of Wigner $D$-functions. Thus, the quantity measured is, 
\begin{equation}\label{eq:wigner_D_expanison}
    w_j = w(\alpha_j,\beta_j,\gamma_j) = \sum_{n,m,\mu}^{n_{\max}} \acoef{n}{m\mu} \WD{n}{\mu}{m}(\alpha_j,\beta_j,\gamma_j) + \eta_j,
\end{equation}
where $j$ indexes the measurement position $(\alpha_j,\beta_j,\gamma_j)\in \sot$, $\eta_j$ is additive measurement noise, and $\WD{n}{\mu}{m}(\alpha,\beta,\gamma)$ are the Wigner $D$-functions with indices satisfying $n\in\{0,1,\cdots, n_{\max}\}$ and $m,\mu \in\{-n,-n+1,\cdots,n-1,n\}$.

The transformations described above result in the series coefficients $\acoef{n}{m\mu}$ being a product of the SW coefficients, the translation transformation, and the probe's receiving coefficients. This product takes the form of $\acoef{n}{m\mu}=\sum_{\nu=\nu_0}^{\nu_{\mrm{max}}}\t{A}{_n^m}\t{B}{_{n,\nu}^{\mu}} \t{R}{_{\nu}^{\mu}}$, where the $\t{A}{_{n}^{m}}$ are the SW coefficients, $\t{B}{_{n,\nu}^{\mu}}$ contains the translation transformation information, $\t{R}{_{\nu}^{\mu}}$ are the receiving coefficients of the probe, and $\nu_0=|\mu| \neq 0$ and $\nu_0=|\mu|$ for EM and acoustics SWs, respectively. If an ideal probe is used, then the receiving coefficients are proportional to the Kronecker delta function. Typically, one chooses a probe as close to ideal as possible and thus, non-ideal probes are electrically small with a significant amount of rotational symmetry, i.e., $\t{R}{_{\nu}^{\mu}} \approx 0$ for $\nu_{\mrm{max}} \gtrsim 10$ and $|\mu| \gtrsim 1$. Consequently, the $\acoef{n}{m\mu}$ will be sparse/compressible if $\t{A}{_n^m}$ coefficients are sparse. Thus, the spherical NF2FF problem is an inverse problem that might be solved using CS. Specifically, if we first recover the Wigner $D$-function coefficients using CS, the SW coefficients can be readily estimated. As a note, even if a slightly non-ideal probe is used, CS should be viable with increases in the number of measurements. This is because the effect will be to moderately increase the sparsity level of the $\acoef{n}{m\mu}$ because non-ideal probes have nonzero receiving coefficients for higher $\mu$.

In the above, the Wigner $D$-function is defined by 
\begin{equation}\label{eq:wigner_D}
    \t{D}{_n^{\mu m}}\left(\alpha,\beta,\gamma\right) = (4\pi^2)^{-1/2}e^{-\mi \mu \alpha}\t{d}{_n^{\mu m}}(\beta) e^{-\mi m \gamma},
\end{equation}
where $\t{d}{_n^{\mu m}}$ is the real Wigner $d$-function defined by
\begin{equation}\label{eq:wigner_d}
    \begin{split}
        \t{d}{_n^{\mu m}}(\beta) =& (-1)^{\mu -m} \sqrt{(2n+1)/{2}} \\
        & \times \sqrt{(n+m)!(n-m)!(n+\mu)!(n-\mu)!} \\
    & \times \sum_{\sigma = \max \left(0, m-\mu\right)}^{\min\left(n+m,n-\mu\right)} \xi_\sigma, \\
    \xi_\sigma = &\frac{
        (-1)^\sigma \left(
            \cos\frac{\beta}{2}\right)
        ^{2n-2\sigma+m-\mu} \left(
            \sin\frac{\beta}{2}
        \right)^{2\sigma-m+\mu}
    }{
        \sigma! (n+m-\sigma)!(n-\mu-\sigma)!(\mu-m+\sigma)!
    }.
    \end{split}
\end{equation}
The normalization factors $(4\pi^2)^{-1/2}$ and $\sqrt{(2n+1)/{2}}$ are chosen for later convenience.

The arguments of the Wigner $D$-function, $(\alpha,\beta,\gamma)\in [0,2\pi)\times[0,\pi]\times[0,2\pi)$, are the Euler angles representing a measurement position on $\mrm{SO}(3)$. The Wigner $D$-functions with $n\leq n_{\max}$ form an orthonormal basis for band-limited functions in $L_2(\mrm{SO}(3))$ and satisfy the orthonormality relation
\begin{equation}\label{eq:normalized_wigner_D_orthogonality}
   \inner{\WD{n}{\mu}{m}}{\WD{n'}{\mu'}{m'}}{\sot} =\t{\delta}{_{nn'}}\t{\delta}{_{mm'}}\t{\delta}{_{\mu\mu'}},
\end{equation} 
where $\t{\delta}{_{nn'}}$ is the Kronecker delta function. Here, $\inner{f}{g}{\sot}$ is the $L_2$ inner product of two functions $f$ and $g$ on $\sot$,
\begin{equation}
    \inner{f}{g}{\sot} = \intg{\mrm{SO}(3)}{}{f(\alpha,\beta,\gamma) \overline{g(\alpha,\beta,\gamma)}}{\sot},
\end{equation}
with $\mrm{d}\mrm{SO}(3)=\sin\beta \mrm{d}\alpha\mrm{d}\gamma \mrm{d}\beta$. The inner product on a subset $R$ of $\sot$ and its compliment $R^\mrm{c}$ are defined as
\begin{equation}
    \inner{f}{g}{R} = \intg{R}{}{f(\alpha,\beta,\gamma) \overline{g(\alpha,\beta,\gamma)}}{\sot}
\end{equation}
and
\begin{equation}
    \inner{f}{g}{R^\mrm{c}} = \intg{R^\mrm{c}}{}{f(\alpha,\beta,\gamma) \overline{g(\alpha,\beta,\gamma)}}{\sot},
\end{equation}
respectively, and will be needed later in the paper. Each of these inner products have an associated norm, $\|f\|_{\sot}^2 = \inner{f}{f}{\sot}$, $\|f\|_R^2=\inner{f}{f}{R}$, and $\|f\|_{R^\mrm{c}}^2=\inner{f}{f}{R^\mrm{c}}$.

It is important for future calculations to note that the band-limited Wigner $D$-functions are uniformly bounded~\cite{bangun_sensing_2020},
\begin{eqnarray}
        \sup_{\substack{ 0\leq n\leq n_{\max} \\  \mu,m \in{-n,\cdots,n}}} \left\|\sqrt{\sin \beta} \WD{n}{\mu}{m}(\alpha,\beta,\gamma)\right\|_\infty & \leq &   C (2n_{\max} + 1)^{1/4} \nonumber \\ 
        & \leq &   C' N_D^{1/12}, \label{eq:Wigner_D_Bound}
\end{eqnarray}
for some constants $C>0$ and $C'>0$. In \eqref{eq:Wigner_D_Bound} $N_D$ as the number of band-limited Wigner $D$-functions with band-limit $n_{\max}$ and is given by
\begin{equation}\label{eq:num_D_funcs}
    N_D=(n_{\max}+1)(2n_{\max}+1)(2n_{\max}+3)/3.
\end{equation}

Casting \eqref{eq:wigner_D_expanison} as \cite{bangun_sensing_2020} does for the CS problem we obtain
\begin{equation} \label{eq:Wigner_D_Inverse_Problem}
    w=\Phi a + \eta,
\end{equation}
where a known set of $M$ measurements $\{w_j\}$ at positions $\{(\alpha_j,\beta_j,\gamma_j)\}$ is denoted by the vector $w$ with entries $w_j$ and the elements of the measurement matrix $\Phi\in \mbb{C}^{M \times N_D}$ are
\begin{equation}\label{eq:wigner_d_measurement_matrix}
    \left[\Phi \right]_{jk} = \WD{n(k)}{\mu(k)}{m(k)}(\alpha_j,\beta_j,\gamma_j)
\end{equation}
with $j\in\{1,2,\cdots,M\}$ and $k\in\{1,2,\cdots,N_D\}$. In \eqref{eq:Wigner_D_Inverse_Problem},
$a\in \mbb{C}^{N_D}$ is the vector of correspondingly ordered coefficients $\acoef{n}{m\mu}$ and $\eta$ is the vector of additive measurement noise. The results in \cite{bangun_sensing_2020} show that if the samples are selected uniformly at random according to, $\alpha_j\sim \mcl{U}(0,2\pi)$, $\beta_j\sim \mcl{U}(0,\pi)$, $\gamma_j\sim \mcl{U}(0,2\pi)$, and the number of measurements, $M$, satisfies
\begin{equation}
    M \geq \widetilde{C} N_D^{1/6} s \ln^3(s) \ln(N_D)
\end{equation}
for some constant $\widetilde{C}>0$, then, after a preconditioning step, standard CS guarantees for BOSs apply. Here, $s$ is the sparsity of the coefficient vector, $a$.

For the inverse problem in \eqref{eq:Wigner_D_Inverse_Problem}, the theoretical guarantee in~\cite{bangun_sensing_2020} requires measurements to be taken at arbitrary positions on all of $\mrm{SO}(3)$. However, many measurement configurations leave parts of $\mrm{SO}(3)$ inaccessible. As an example, certain measurement systems have the antenna placed on a support structure with its main lobe pointing up. In such a setup, any values of $\alpha$ and $\gamma$ are available to measure, but certain values or intervals of $\beta$ are not. Our work in Section \ref{sec:cs_on_R} shows that such restrictions on measurement positions do not invalidate the use of QCBP and its associated CS guarantees. However, restrictions do invalidate the bound on the number of measurements needed for reconstruction in~\cite{bangun_sensing_2020}. This invalidation is because restricting measurement positions prohibits uniform sampling of the measure associated with the preconditioned Wigner $D$-function BOS.  As a note, \cite{bangun_sensing_2020} and other similar papers, e.g.~\cite{hofmann_minimum_2019}, also perform CS by testing with specific sampling grids or by minimizing the coherence of the measurement matrix. Such methods can be used with a restriction to measurements from $R$, but we emphasize that these methods still would not provide the theoretical guarantees we seek. 

Generally, we can think of the restriction above as limiting $\beta$ to a range from $\Theta_1$ to $\Theta_2$. This is sometimes known as a latitudinal belt when considering a sphere instead of $\mrm{SO}(3)$, but we will use the same name. The latitudinal belt is the particular form of restricted measurements we consider in this paper. We show that restriction to a latitudinal belt requires a change of the basis from the Wigner-$D$ functions to a related set of Slepian functions on $\mrm{SO}(3)$. This change of basis does increase the coefficient sparsity level slightly, but it does so in a bounded way by only transforming within special fixed subspaces of Wigner $D$-functions. Although the transformation allows us to show that CS guarantees hold, the required number of measurements increases by a factor proportional to the square of the dimension of the largest fixed subspace.

\subsection{Slepian Functions on the Rotation Group} \label{sec:rgsfs}

\subsubsection{Concentration on a general sub-domain}
The spectral concentration problem gives rise to Slepian functions on a domain \cite{slepian_prolate_1961, landau_prolate_1961, landau_prolate_1962, simons_scalar_2015}. These special functions are an orthogonal basis for band-limited $L_2$ functions on their entire domain and on the subset of the domain to which they are space-limited. This property is of particular interest in our setting because measurements of the field are not on all of $\mrm{SO}(3)$ but instead on a subset of $\mrm{SO}(3)$. 

Though frequently constructed on the domain $\mbb{R}$, recent work has shown that Slepian functions can be constructed on more complex domains like the sphere~\cite{simons_scalar_2015}. This work showed that using a basis for $L_2$ functions on the entire domain, one can construct Slepian functions concentrated on a chosen sub-domain. Using this approach, we develop Slepian functions concentrated on a latitudinal belt on $\mrm{SO}(3)$ using only Wigner $D$-functions. Mathematically, we formulate a matrix problem that maximally concentrates linear combinations of band-limited Wigner $D$-functions on $R\subseteq\mrm{SO}(3)$. We call the resulting functions the Rotation Group Slepian Functions (RGSFs) and denote them by $\t{g}{_1}, \t{g}{_2}, \dots, \t{g}{_N} \in L_2(\mrm{SO}(3))$. Importantly, the RGSFs can be shown to satisfy a uniform bound that is useful for CS recovery.

The requirements of band-limitedness and $L_2$ integrability imply that each of the RGSFs has an expansion of the form 
\begin{equation}\label{eq:rg_slepian_expansion}
    \t{g}{_i}(\alpha, \beta, \gamma) = \sum_{n,m,\mu}^{n_{\max}} \t{\widetilde{g}}{_{in}^{\mu m}} \WD{n}{\mu}{m}(\alpha, \beta, \gamma),
\end{equation}
where the coefficients $\t{\widetilde{g}}{_{in}^{\mu m}}$ are unknown. To find these coefficients (subject to orthogonality with all $\t{g}{_j}$, $j<i$), the following optimization problem is solved iteratively over the index $i$:
\begin{equation}\label{eq:maximization_integral_eq}
    \begin{split}
        &\maximize_{\t{\widetilde{g}}{_{in}^{\mu m}}\in\mathbb{C}} \lambda_i \\ 
        &\st \begin{cases}
            \lambda_i = \frac{ \|\t{g}{_i}\|_R^2
            }{
                \|\t{g}{_i}\|_{\sot}^2
            }, \\
            0 = \inner{\t{g}{_i}}{\t{g}{_j}}{\sot}, \; \forall j<i \leq N_D.
        \end{cases}
    \end{split}
\end{equation}
This optimization problem can be viewed as maximizing the concentration of $\t{g}{_i}$ on $R$ relative to all of $\mrm{SO}(3)$ while ensuring the entire set of $\t{g}{_i}$ are mutually orthogonal on $\mrm{SO}(3)$. Substituting \eqref{eq:rg_slepian_expansion} into \eqref{eq:maximization_integral_eq} and using \eqref{eq:normalized_wigner_D_orthogonality} we find
\begin{equation}\label{eq:Slepian_eigenvalues}
    \lambda_i = \frac{
        \sum_{n',\mu',m'}^{n_{\max}} \overline{\t{\widetilde{g}}{_{in'}^{\mu' m'}}} \sum_{n,\mu,m}^{n_{\max}} \inner{\WD{n}{\mu}{m}}{\WD{n'}{\mu'}{m'}}{R} \t{\widetilde{g}}{_{in}^{\mu m}}
    }{
        \sum_{n,\mu,m}^{n_{\max}} \overline{\t{\widetilde{g}}{_{in}^{\mu m}}}  \t{\widetilde{g}}{_{in}^{\mu m}}
    }.
\end{equation}
We recognize \eqref{eq:Slepian_eigenvalues} as the equivalent matrix equation
\begin{equation}
    \lambda_i = \frac{\mbf{g}_i^*\msc{D}\mbf{g}_i}{\mbf{g}_i^*\mbf{g}_i}
\end{equation}
where the elements of the matrix $\msc{D}\in \mathbb{C}^{N_D \times N_D}$ are inner products between Wigner $D$-functions on the domain $R$, 
\begin{equation}\label{eq:rgsf_matrix}
    [\msc{D}]_{jk} =  \inner{\WD{n(k)}{\mu(k)}{m(k)}}{\WD{n'(j)}{\mu'(j)}{m'(j)}}{R},
\end{equation}
and the vector $\mbf{g}_i$ has the coefficients $\t{\widetilde{g}}{_{in}^{\mu m}}$ correspondingly arranged according to the ordering functions $n(j),\; \mu(j),\; m(j)$. Furthermore, iteratively solving \eqref{eq:maximization_integral_eq} is equivalent to iteratively solving the constrained maximum eigenvalue problem
\begin{equation}\label{eq:general_eval_problem}
    \maximize_{\mbf{g}_i \in\mathbb{C}^{N_D}} \lambda_i \quad \st \begin{cases}
        \lambda_i \mbf{g}_i = \msc{D}\mbf{g}_i, \\
        0 = \mbf{g}_j^*\mbf{g}_i,\; \forall j<i \leq N_D.
    \end{cases}
\end{equation}

From \eqref{eq:rgsf_matrix} and \eqref{eq:maximization_integral_eq} $\msc{D}$ is Hermitian and positive definite, respectively. This implies $\msc{D}$ has a spectral decomposition with orthonormal eigenvectors and, by construction, the associated eigenvalues satisfy $1> \lambda_1\geq\lambda_2 \geq \cdots \lambda_{N_D} > 0$. It follows that the solution \eqref{eq:maximization_integral_eq}, and thus the set of RGSFs, is given by the eigenvalues and associated eigenvectors of $\msc{D}$. Importantly, the RGSFs are orthogonal over both $\mrm{SO}(3)$ and $R$ (see \Cref{sec:ortho_derivation} for the derivation), i.e.,
\begin{equation} \label{eq:SO3_orthogonality}
    \inner{\t{g}{_{i}}}{\t{g}{_{i'}}}{\sot} = \delta_{ii'}
\end{equation}
and
\begin{equation} \label{eq:R_orthogonality}
    \inner{\t{g}{_{i}}}{\t{g}{_{i'}}}{R} = \lambda_i\delta_{ii'}.
\end{equation}

The above derivation is general and works for any region $R$. Before specializing the choice of $R$ to a latitudinal belt, we make several remarks.

\begin{remark}
    As shown in \eqref{eq:maximization_integral_eq}, the eigenvalue associated with each RGSF measures the degree to which it is localized to $R$. So we expect the RGSFs with eigenvalues close to unity to be well localized on $R$ and the RGSFs with eigenvalues near zero to be more localized on $R^\mrm{c}$.
\end{remark}

\begin{remark}
    The total number of RGSFs is always $N_D$ and does not depend on $R$. However, the number of RGSFs highly concentrated on $R$ (i.e., $\lambda_i$ near unity) does depend on the selection of $R$.
\end{remark}

\begin{remark}
    Similar to how we can expand the RGSFs in Wigner $D$-functions, we can expand the Wigner $D$-functions in the RGSFs. This is because both sets of functions form a basis for band-limited functions in $L_2(\mrm{SO}(3))$.
\end{remark}

\subsubsection{Concentration on a latitudinal belt}

We are interested in the case when measurements are limited to a latitudinal belt, so we set the restricted region to be $R=[0,2\pi)\times[\Theta_1,\Theta_2]\times[0,2\pi)$ with $0\leq \Theta_1 < \Theta_2 \leq \pi$. Under this restriction, using \eqref{eq:wigner_D}, and since the $\Wd{n}{\mu}{m}$ are real, the elements of the matrix $\msc{D}$ reduce to 
\begin{equation}\label{eq:das_rgsf_matrix}
    \begin{split}
        [\msc{D}]_{jk} =& \;\delta_{m(k)m(j)}\delta_{\mu(k) \mu(j)} \\ 
        & \times \intg{\Theta_1}{\Theta_2}{
            \Wd{n(k)}{\mu(k)}{m(k)}(\beta) \Wd{n(j)}{\mu(j)}{m(j)}(\beta) \sin\beta
        }{\beta}.
    \end{split}
\end{equation}
Thus, there exists an indexing such that $\msc{D}$ has a block diagonal structure,
\begin{equation}
    \msc{D} = \mrm{diag}\left(\t{\msc{D}}{^{0 0}},\t{\msc{D}}{^{0 1}},\t{\msc{D}}{^{0 -1}},\t{\msc{D}}{^{10}}, \cdots,  \t{\msc{D}}{^{-n_{\max} - n_{\max}}}\right)
\end{equation}
with
\begin{equation}
    [\t{\msc{D}}{^{\mu m}}]_{jk} = \intg{\Theta_1}{\Theta_2}{
        \Wd{n(k)}{\mu}{m}(\beta) \Wd{n(j)}{\mu}{m}(\beta) \sin\beta
    }{\beta}.
\end{equation}
The matrix $\msc{D}$ will have a total of $(2n_{\max}+1)^2$ blocks and each block has a dimension of $n_{\max}-n_{\min}+1$, where $n_{\min}=\max(|m|,|\mu|)$. 

From \eqref{eq:das_rgsf_matrix}, the block diagonal structure arises due to the presence of the full $\alpha$ and $\gamma$ ranges. This makes $[\msc{D}]_{jk}$ zero unless the matrix pairs functions that belong to the same fixed $\mu$, $m$ subspace $S^{m\mu}$. Specifically, $S^{m\mu}$ can be defined as,
\begin{equation}
    S^{m\mu} = \mrm{span} \left\{ \WD{n}{\mu}{m}(\alpha, \beta, \gamma): n\in \{n_{\min},\cdots,n_{\max}\} \right\}, 
\end{equation}
where $m,\;\mu\in\{-n_{\max},-n_{\max}+1,\cdots,n_{\max}\}$.

Importantly, each of the sub-matrices $\t{\msc{D}}{^{\mu m}}$ are symmetric and can be solved individually for their eigenvectors. For convenience and to maintain awareness of the subspace $S^{m\mu}$, we change notation and label the eigenvectors with their associated $m,\;\mu$ values and order them according to decreasing eigenvalues within $S^{m\mu}$. That is, the eigenvectors and eigenvalues are $\t{\mbf{g}}{_i^{\mu m}}$ and $\lambda_{i}^{\mu m}$, respectively, where $i\in\{1,\cdots,n_{\max}-n_{\min}+1\}$, $\mu,\; m\in\{-n_{\max},-n_{\max}+1,\cdots,n_{\max}\}$, and $\lambda_{1}^{\mu m}\geq \lambda_{2}^{\mu m} \geq \cdots \geq \lambda_{n_{\max}-n_{\min}+1}^{\mu m}$. To relate the two labeling schemes, if we take all $\t{\mbf{g}}{_i^{\mu m}}$ and order them according to decreasing eigenvalues, we get the original labeling $\t{\mbf{g}}{_i}$ with $i$ ranging from $1$ to $N_D$.  

The eigenvectors of each $\t{\msc{D}}{^{\mu m}}$ give the expansion coefficients for the RGSFs in $S^{m\mu}$. Similar to the eigenvectors, we will re-index the RGSFs by $m$ and $\mu$ and their order in concentration within $S^{m\mu}$. Since the eigenvectors $\t{\mbf{g}}{_i^{\mu m}}$ only involve the fixed $S^{m\mu}$ subspace we can rewrite~\eqref{eq:rg_slepian_expansion} as
\begin{equation}\label{eq:das_rgsf_expansion}
    \t{g}{_{i}^{\mu m}}(\alpha, \beta, \gamma) = \sum_{n=n_{\min}}^{n_{\max}}\t{\widetilde{g}}{_{in}^{\mu m}}\WD{n}{\mu}{m}(\alpha, \beta, \gamma)
\end{equation}
with the inverse relation given by (see supplementary material \Cref{sec:inverse_derivation} for derivation)
\begin{equation}\label{eq:inverse_rgsf_expansion}
    \WD{n}{\mu}{m}(\alpha, \beta, \gamma) = \sum_{i=1}^{n_{\max}-n_{\min} +1} \t{\widetilde{g}}{_{in}^{\mu m}} \t{g}{_{i}^{\mu m}}(\alpha, \beta, \gamma),
\end{equation}
where $i\in\{1,\cdots,n_{\max}-n_{\min}+1\}$ and $m,\;\mu\in\{-n_{\max},-n_{\max}+1,\cdots,n_{\max}\}$. Importantly, due to the maximum dimension of any $S^{m\mu}$, the RGSFs satisfy the following uniform bound.

\begin{proposition} \label{prop:rgsf_bound}
    Let $R$ be the latitudinal belt defined by $\Theta_1$ and $\Theta_2$, $\t{g}{_{i}^{\mu m}}(\alpha,\beta,\gamma)$ be the RGSF on $R$ associated with the $i$th eigenvalue of the matrix $\t{\msc{D}}{^{\mu m}}$ corresponding to the subspace indicated by $m$ and $\mu$, $\mcl{S}$ be the subset of band-limited RGSFs on $R$ with eigenvalues greater than or equal to $\lambda_\text{min}$, and $N_D$ be the total number of band-limited RGSFs on $R$. Then the following holds:
    \begin{equation}\label{eq:RGSF_Bound}
        \sup_{\t{g}{_{i}^{\mu m}}\in \mcl{S}} \left\| \sqrt{\frac{\sin \beta}{\t{\lambda}{_i^{\mu m}}}} \t{g}{_i^{\mu m}}(\alpha, \beta, \gamma)\right\|_\infty \leq \frac{C'' N_D^{1/4}}{ \sqrt{ \lambda_\text{min} } } 
    \end{equation}
    where $C'' = \left(\frac{3}{2}\right)^{1/6}C'$ and $C'>0$ is from \eqref{eq:Wigner_D_Bound}.
\end{proposition}

\tit{Proof}: See supplementary material \Cref{sec:bound_prop_proof}.

The construction of RGSFs on the latitudinal belt enables CS recovery to remain viable on $R$. The particular selection of $R$ is such that the $\t{\msc{D}}{^{\mu m}}$ result in isolated rotations of Wigner $D$-functions within each subspace $S^{m\mu}$. Moreover, these subspaces have a maximum dimension $n_{\max}+1$. Therefore, even without explicitly knowing the eigenvectors, their transformation does not grow the bound on the RGSFs too large, and thus the construction is useful for CS recovery. This is formalized in Proposition~\ref{prop:rgsf_bound}.

\section{Compressive Sensing Guarantees for Structured Matrices} \label{sec:cs_stuff}

Compressive sensing approaches to the inverse problems presented in this paper rely on two key assumptions. First is that the unknown vector of coefficients is sparse (having many zero entries) or be compressible. Roughly speaking, compressible means that a vector is well approximated by a sparse vector. Work in \cite{loschenbrand_fast_2016, culotta-lopez_compressed_2018} has established that this is indeed the case in the Wigner $D$-function basis for many antennas of interest. The second assumption concerns the choice of the measurement matrix $\Phi$, which must allow for efficient recovery of a sparse or compressible vector via QCBP. To guarantee the recovery performance of QCBP, we argue that the types of $\Phi$ we consider satisfy the RIP with high probability. The relevant background on the RIP and performance guarantees are given in the remainder of this section.

\begin{definition}[Best $s$-Sparse Approximation Error~{\cite[p.~42, def.~2.2]{foucart_mathematical_2013}}]
    Given a vector $x\in\mathbb{C}^N$, the best $s$-sparse approximation error in the $\ell_p$ norm is
    \begin{equation}
        \sigma_s(x)_p = \inf_{z\in\mathbb{C}^N:\|z\|_0\leq s} \|z-x\|_p.
    \end{equation}
\end{definition}

\begin{definition}[Restricted Isometry Property (RIP)~{\cite[p.~133, def.~6.1]{foucart_mathematical_2013}}]
    A matrix $\Phi\in \mathbb{C}^{M\times N}$ satisfies the restricted isometry property of order $s$ with constant $\delta \in (0,1)$ if the following holds for all $s$-sparse vectors in $x\in \mathbb{C}^N$
    \begin{equation}
        (1-\delta) \|x\|_2^2\leq \|\Phi x\|_2^2\leq (1+\delta) \|x\|_2^2,
    \end{equation}
    where the smallest $\delta$, denoted by $\delta_s$, is called the restricted isometry constant.
\end{definition}

\begin{theorem}[RIP for Bounded Orthonormal Systems (BOS)~{\cite[p.~405, Thm.~12.31]{foucart_mathematical_2013}}]\label{thm:rip_bos}
    Consider a set of bounded orthonormal basis functions $\phi_i:\mcl{D}\rightarrow \mathbb{C},\; i\in\{1,2,\cdots,N\}$ that are orthonormal with respect to a probability measure $\rho$ on the measurable space $\mcl{D}$. Consider the matrix $\Phi\in \mathbb{C}^{M \times N}$ with entries 
    \begin{equation}
        \phi_{ji} = \phi_i(t_j),\; j\in\{1,2,\cdots,M\},\; i\in\{1,2,\cdots,N\}
    \end{equation}
    constructed with i.i.d. samples of $t_j$ from the measure $\rho$ on $\mcl{D}$. Suppose the orthonormal functions are bounded such that $1\leq\sup_{i\in\{1,\cdots,N\}} \|\phi_i\|_\infty\leq K$. If
    \begin{equation}
        M \geq C_0 \delta^{-2} K^2 s \ln^4(N)
    \end{equation}
    then with a probability of at least $1-N^{-\ln^3(N)}$, the restricted isometry constant $\delta_s$ of $\frac{1}{\sqrt{M}}\Phi$ satisfies $\delta_s\leq \delta$ for $\delta \in(0,1)$. The constant $C_0>0$ is universal.
\end{theorem}

\begin{theorem}[Sparse Recovery for RIP Matrices~{\cite[p.~144, Thm.~6.12]{foucart_mathematical_2013}}] \label{thm:sparse_recovery_rip}
    Suppose that the matrix $\Phi\in\mathbb{C}^{M\times N}$ has restricted isometry constant $\delta_{2s} < 4/\sqrt{41} \approx 0.6246$. Suppose that the measurements are taken with $\Phi$ and are noisy, $y=\Phi x +\eta$, with $\|\eta\|_\infty\leq \epsilon$. If $\widehat{x}$ is a solution to 
    \begin{equation}
        \widehat{x} = \arg \min_{z\in\mathbb{C}^N}  \|z\|_1 \; \text{subject to} \; \|y-\Phi z\|_2 \leq \sqrt{M}\epsilon,
    \end{equation}
    then 
    \begin{equation}
        \|x-\widehat{x}\|_2 \leq C_1\left( \frac{\sigma_s(x)_1}{\sqrt{s}} + \epsilon \right),
    \end{equation}
    where the constant $C_1>0$ only depends on $\delta_{2s}$.
\end{theorem}

Theorem \ref{thm:sparse_recovery_rip} states that a compressible vector $x$ can be recovered using QCBP with an accuracy proportional to the sparse approximation error plus a noise factor so long as the measurement matrix $\Phi$ satisfies RIP with small enough RIP constant. Pairing Theorem \ref{thm:rip_bos} and Theorem \ref{thm:sparse_recovery_rip} shows that BOS measurement matrices can be used for accurate CS recovery provided that enough samples are taken.

\section{Compressive Sensing on the Restricted Domain $R$} \label{sec:main_res}
The approach we take to achieve our main result in Theorem \ref{thm:restricted_so3_accuracy} is broken into three parts. In Section \ref{sec:transform_WD_to_rgsf}, we transform the measurement matrix $\Phi$ and the coefficient vector $a$ in \eqref{eq:Wigner_D_Inverse_Problem} from the Wigner $D$-function to the RGSF basis. We then state our main result in Section~\ref{sec:cs_on_R} along with our interpretation of this result. In Section~\ref{sec:transformation_of_sparsity} we detail the effect of the transformation presented in Section~\ref{sec:transform_WD_to_rgsf} on the sparsity of the coefficient vector $a$. In particular, we argue that for certain cases, the sparsity level could increase, but not so much that the coefficient vector in the RGSF basis is no longer sparse, and this can be determined from the sparsity pattern of the Wigner $D$-function coefficients only.

\subsection{Transformation of the Inverse Problem in \eqref{eq:Wigner_D_Inverse_Problem}}\label{sec:transform_WD_to_rgsf}
Using~\eqref{eq:wigner_d_measurement_matrix} and~\eqref{eq:inverse_rgsf_expansion}, we can write the term $\Phi a$ appearing in~\eqref{eq:Wigner_D_Inverse_Problem} as 
\begin{equation}\label{eq:phia-principle-sqrt}
    \Phi a = \Phi U^* \sqrt{\Lambda}^{-1} \sqrt{\Lambda} U a = \Phi' \sqrt{\Lambda} U a, 
\end{equation}
where $\sqrt{\Lambda}\in \mbb{C}^{M\times N_D}$ is the principal square root of the diagonal matrix of strictly positive RGSF concentrations, i.e.,
\begin{equation}
    \left[\Lambda\right]_{jk} = \t{\lambda}{_{i(j)}^{\mu(j) m(j)}}\t{\delta}{_{jk}}.
\end{equation}
In \eqref{eq:phia-principle-sqrt}, $U\in \mbb{C}^{M\times N_D}$ is the unitary matrix that transforms from the Wigner $D$-function basis into the RGSF basis (see \eqref{eq:das_rgsf_expansion}) with elements
\begin{equation}
    \left[ U\right]_{jk} = \t{\widetilde{g}}{_{i(j)n(k)}^{\mu(j)m(j)}}
\end{equation}
and $\Phi' = \Phi U^* \sqrt{\Lambda}^{-1}\in \mbb{C}^{M\times N_D}$ has elements
\begin{equation}
    \left[\Phi'\right]_{jk} = \left(\t{\lambda}{_{i(k)}^{\mu(k) m(k)}}\right)^{-1/2} \t{g}{_{i(k)}^{\mu(k) m(k)}}(\alpha_j,\beta_j,\gamma_j).
\end{equation}

Next, we absorb $\sqrt{\Lambda}U$ into the unknown vector of coefficients, giving
\begin{equation}
    \Phi a = \Phi' \sqrt{\Lambda} U a = \Phi' \sqrt{\Lambda} \widetilde{a} = \Phi' a', 
\end{equation}
where $\widetilde{a} = U a$ is the vector of coefficients in the RGSF basis normalized over all of $\mrm{SO}(3)$ and $a' = \sqrt{\Lambda} \widetilde{a}$ is the vector of coefficients on the RGSFs normalized on $R$. We now write our inverse problem as
\begin{equation} \label{eq:rotated_matrix_problem}
    w = \Phi' a' + \eta.
\end{equation}
From the form of $U$ that derives from \eqref{eq:das_rgsf_expansion} and \eqref{eq:inverse_rgsf_expansion}, if the signal $a$ is sparse or compressible due to a small subset of all $S^{m\mu}$ being sufficient in the Wigner $D$-function basis, then it will remain so in the RGSF basis. This is detailed in Section~\ref{sec:transformation_of_sparsity}.

The inverse problem in \eqref{eq:rotated_matrix_problem} is a manipulated version of \eqref{eq:Wigner_D_Inverse_Problem} and no assumptions about the set of measurement positions have been made. However, it is more amenable to choosing measurement positions that are restricted to $R$. A drawback, however, is that some $\lambda_i^{\mu m}$ are very small and so the $R$-normalized RGSFs' uniform bound will be very large.

To avoid the problem caused by RGSFs with small $\lambda_i^{\mu m}$, we take the columns of $\Phi'$ with the RGSFs having eigenvalues less than some cutoff $\lambda_c$, say, $N_{\lambda<\lambda_c}$ functions, and place them in a matrix $\Phi_2'$. We also take the remainder of the columns to be in a matrix $\Phi_1'$ and partition $a'$ accordingly into $a_1'$ and $a_2'$. This allows us to write \eqref{eq:rotated_matrix_problem} as
\begin{equation}\label{eq:partitioned_matrix_problem_2}
    w = \Phi_1'a_1' + \Phi_2'a_2' + \eta.
\end{equation}
Without detailing the effect here (see Section \ref{sec:cs_on_R}), it is beneficial to group the last two terms above into the new noise term $\eta'$ to obtain
\begin{equation} \label{eq:transformed_truncated_matrix_problem}
    w = \Phi_1'a_1' + \eta'.
\end{equation}
With an eye towards \eqref{eq:RGSF_Bound}, we precondition \eqref{eq:transformed_truncated_matrix_problem} with the diagonal matrix $P_{ii} = \sqrt{\sin \beta_i}$. Then, $P\Phi_1'$ is a matrix constructed from a sampled BOS over the latitudinal belt $R$. In particular, $P\Phi_1'$ arises by sampling the BOS containing only preconditioned RGSFs with concentrations no less than $\lambda_c$. Explicitly, the BOS is the set $\mcl{S}$ with $\lambda_{\min}=\lambda_c$, 
\begin{equation}\label{eq:RGSF_BOS}
    \mcl{S}' \coloneqq \left\{\sqrt{\frac{\sin\beta}{\t{\lambda}{_i^{\mu m}}}} \t{g}{_{i}^{\mu m}}(\alpha, \beta, \gamma):\; \t{g}{_{i}^{\mu m}} \in \mcl{S} \text{ with } \lambda_\text{min}=\lambda_c\right\}
\end{equation}
paired with the measure $\mrm{d}\rho=\mrm{d}\alpha \mrm{d}\beta \mrm{d}\gamma$ over the domain $R$. In this preconditioned formulation we can readily apply the CS results from Section~\ref{sec:cs_stuff} to solve for $a_1'$. 

To summarize, we took~\eqref{eq:Wigner_D_Inverse_Problem} with a measurement matrix constructed from the BOS of band-limited Wigner $D$-functions and transformed it into~\eqref{eq:transformed_truncated_matrix_problem} with a measurement matrix constructed from the BOS of band-limited RGSFs with eigenvalues no less than $\lambda_c$. In \eqref{eq:Wigner_D_Inverse_Problem}, we solve for the coefficient vector $a$, requiring measurements on all of $\sot$ for CS to apply. In~\eqref{eq:transformed_truncated_matrix_problem}, we solve for the coefficient vector $a_1'$, requiring measurements only on $R$. Here $a_1'$ contains the coefficients of the function $w(\alpha,\beta,\gamma)$ on the RGSFs with concentrations $\lambda_c$ or greater. Therefore, solving \eqref{eq:transformed_truncated_matrix_problem} will not give all of the RGSF coefficients for $w(\alpha,\beta,\gamma)$ and therefore cannot be expected to give trustworthy field values on $R^{\mrm{c}}$ where the unknown coefficients in $a_2'$ contribute to $w(\alpha,\beta,\gamma)$. However, \Cref{sec:cs_on_R} will show that having only $a_1'$ does not prohibit the accurate recovery of $w(\alpha,\beta,\gamma)$ if $w$ falls into a certain class of functions. 

\subsection{Compressive Sensing on $R$} \label{sec:cs_on_R}
We now state our main result for applying CS on the latitudinal belt $R$ by solving \eqref{eq:transformed_truncated_matrix_problem}. This is given in Theorem \ref{thm:restricted_so3_accuracy} below. Note, the last part of this result, \eqref{eq:b_bound}, characterizes the size of the truncated RGSF coefficients $a_2'$ in terms of the energy of $w$ on $R^{\mrm{c}}$. In the next subsection, we provide a set of lemmas that are helpful in establishing the sparsity of $a'$ and thus $a_1'$ appearing in~\eqref{eq:transformed_truncated_matrix_problem}.

First, we recap the definitions needed for our main result:
\begin{enumerate}
    \item $a\in \mathbb{C}^{N_D}$ is the vector containing the Wigner $D$-function coefficients for the band-limited function $w(\alpha, \beta, \gamma) = \sum_{n,m,\mu}^{n_{\max}} \acoef{n}{m\mu} \WD{n}{\mu}{m}(\alpha,\beta,\gamma)$ and $N_D$ is the number of band-limited Wigner $D$-functions with band-limit $n_{\max}$.
    \item $R$ is the latitudinal belt on $\sot$ specified by $\Theta_1$ and $\Theta_2$ where $0\leq\Theta_1<\Theta_2\leq \pi$.
    \item The set $\{(\t{\lambda}{_{i(k)}^{\mu(k) m(k)}})^{-1/2} \t{g}{_{i(k)}^{\mu(k) m(k)}}\}_{k=1}^{N_D}$ is the band-limited $R$-normalized RGSF basis for $\sot$ concentrated on $R$.
    \item $a' \in \mathbb{C}^{N_D}$ is the vector of RGSF coefficients for $w(\alpha, \beta, \gamma)$ in the $R$-normalized RGSF basis. The product matrix $\sqrt{\Lambda} U$ transforms the band-limited Wigner $D$-function coefficients $a$ into the band-limited $R$-normalized RGSF coefficients $a'$.
    \item $a_1'$ is the vector of coefficients for $w(\alpha, \beta, \gamma)$ associated with the RGSFs that have $\lambda_i^{\mu m}\geq \lambda_c$, with $\lambda_c \in(\min \lambda_i^{\mu m}, \max \lambda_i^{\mu m})$, and $a_2'$ is the vector of the remaining RGSF coefficients for $w(\alpha, \beta, \gamma)$.
    \item $N_{\lambda<\lambda_c}$ is the number of band-limited RGSFs with concentrations less than $\lambda_c$,
    \item $\Phi_i'$, $i\in\{1,2\}$ is the sensing matrix whose elements are the RGSFs corresponding to $a_i'$. These have elements 
    \begin{equation*}
    [\Phi_i']_{jk} = \frac{\t{g}{_{i(k)}^{\mu(k) m(k)}}(\alpha_j,\beta_j,\gamma_j)}{\sqrt{\t{\lambda}{_{i(k)}^{\mu(k) m(k)}}}}
    \end{equation*}
    for RGSFs satisfying $\t{\lambda}{_{i(k)}^{\mu(k) m(k)}}\geq \lambda_c$ if $i=1$ and $\t{\lambda}{_{i(k)}^{\mu(k) m(k)}}< \lambda_c$ if $i=2$.
    \item $\eta' = \Phi_2'a_2'+\eta$ where $\eta$ is the vector of additive measurement noise.
    \item  $P$ is the diagonal matrix whose elements are $P_{jj} = \sqrt{\sin \beta_j}$.
\end{enumerate}

\begin{theorem}[Sparse Recovery for Wigner $D$-Function Series on Restricted $\mrm{SO(3)}$] \label{thm:restricted_so3_accuracy}
    Consider the linear inverse problem,
    \begin{equation}\label{eq:preconditioned_rgsf_inverse_problem}
        P w = P \Phi_1' a_1' + P \eta',
    \end{equation}
    that is constructed from $M$ i.i.d.\ noise corrupted samples $(\alpha_j,\beta_j,\gamma_j)$ of $w(\alpha,\beta,\gamma)$ where $\alpha_j\sim \mcl{U}(0,2\pi)$, $\beta_j\sim \mcl{U}(\Theta_1,\Theta_2)$, $\gamma_j\sim \mcl{U}(0,2\pi)$, Suppose $\|P\eta'\|_\infty\leq \epsilon$ and $s$ is an integer satisfying
    \begin{equation}
        M\geq \frac{C_2 \sqrt{N_D} }{\lambda_{c}} s \ln^4(N_D).
    \end{equation}  
    Let the estimated RGSF coefficients $\widehat{a}'$ be constructed by solving \eqref{eq:preconditioned_rgsf_inverse_problem} using QCBP for $\widehat{a}_1'$ and setting $\widehat{a}_2'$ to be all zeros. Let the estimated Wigner $D$-function coefficients be $\widehat{a}=U^*\sqrt{\Lambda}^{-1} \widehat{a}'$ so that we have the estimated function $\widehat{w}(\alpha, \beta, \gamma)= \sum_{n,m,\mu}^{n_{\max}} \ahcoef{n}{m\mu} \WD{n}{\mu}{m}(\alpha,\beta,\gamma)$ on $\sot$. Then with probability $1-(N_D-N_{\lambda<\lambda_c})^{-\ln^3(N_D-N_{\lambda<\lambda_c})}$, the following hold.
    \begin{enumerate}[(i)]
        \item The estimated function $\widehat{w}(\alpha, \beta, \gamma)$ satisfies 
        \begin{equation}\label{eq:function_error_on_R}
            \begin{split}
                \|w(\alpha, &\beta, \gamma) - \widehat{w}(\alpha, \beta, \gamma)\|_R^2  \\
                &<C_1^2\left( \frac{\sigma_s(a_1')_1}{\sqrt{s}} + \epsilon \right)^2 + \frac{\lambda_c}{1-\lambda_c} E_{R^\mrm{c}},
            \end{split}
        \end{equation}
        where $E_{R^\mrm{c}}=\|w(\alpha,\beta,\gamma)\|_{R^\mrm{c}}^2$ is the energy of $w(\alpha,\beta,\gamma)$ on $R^\mrm{c}$.
        \item The estimated Wigner $D$-function coefficients $\widehat{a}$ satisfy
        \begin{equation} \label{eq:D_coef_error}
            \begin{split}
                \|a-\widehat{a}\|^2 \leq & \; C_1^2\left( \frac{\sigma_s(a_1')_1}{\sqrt{s}} + \epsilon \right)^2 + \frac{1}{1-\lambda_c}E_{R^\mrm{c}} \\
                & + \widehat{E}_{R^\mrm{c}} + 2\sqrt{E_{R^\mrm{c}} \widehat{E}_{R^\mrm{c}}},
            \end{split}
        \end{equation}
        where $\widehat{E}_{R^\mrm{c}}$ is the energy of $\widehat{w}(\alpha, \beta, \gamma)$ on $R^\mrm{c}$.
        \item The noise term $\|P\eta'\|_\infty \leq \epsilon$ in the errors is bounded, $\|P\eta'\|_\infty \leq \|P\eta\|_\infty + \|P\Phi_2'a_2'\|_\infty$ and $\|P\Phi_2'a_2'\|_\infty$ satisfies
        \begin{equation} \label{eq:b_bound}
            \|P\Phi_2'a_2'\|_\infty \leq \frac{C'' N_D^{3/4} E_{R^\mrm{c}}}{(1-\lambda_c)}.
        \end{equation}
    \end{enumerate}

\end{theorem}

\tit{Proof}: See supplementary material Section VIII-D.

\begin{remark}
    Relationships between the various constants in the above equations are as follows. We have $C_2 = C'' C_0 \delta_{2s}^{-2}$, where $C_0$ is from \Cref{thm:rip_bos} (see associated reference for numerical value) and $C'' = (\frac{3}{2})^{1/6} C'$ with $C'$ found in \eqref{eq:Wigner_D_Bound} (see associated reference for numerical value). Remaining is $C_1$, which is from \Cref{thm:sparse_recovery_rip} (see associated reference for numerical value).
\end{remark}

\begin{remark}
    In the above, the $s$ can be thought of as the sparsity of the vector $a_1'$, which is bounded by the sparsity of $a'$. Thus, Theorem~\ref{thm:restricted_so3_accuracy} tells us that if the sparsity of $a'$ ($a_1'$) scales slower than the square root of $N_D$, the required number of measurements to recover $w(\alpha,\beta,\gamma)$ on $R$ only becomes sub-linear in $N_D$ (ignoring log factors) and the errors in  $w(\alpha,\beta,\gamma)$ and the Wigner $D$-function coefficients are given by \eqref{eq:function_error_on_R} and \eqref{eq:D_coef_error}, respectively.
\end{remark}

\begin{remark}
    From \eqref{eq:function_error_on_R} as $\lambda_c$ decreases, the error in $\widehat{w}(\alpha, \beta, \gamma)$ on $R$ becomes bounded only by the sparse representation error and $\epsilon$, which includes noise and the ignored RGSFs. Also note that if the function $w$ can be represented by only the RGSFs that are kept in $\Phi_1'$, then the bound on function reconstruction error in \eqref{eq:function_error_on_R} depends only on the sparse representation and noise errors (see the proof of \eqref{eq:function_error_on_R}).
\end{remark}

\begin{remark}
    It is worth commenting on the presence of $\widehat{E}_{R^\mrm{c}}$ in \eqref{eq:D_coef_error}. This term comes about because no samples of the function $w(\alpha,\beta,\gamma)$ on $R^\mrm{c}$ are used. In particular, we can write the $l_2$ error for the Wigner $D$-function coefficients in terms of \eqref{eq:function_error_on_R} plus the contributions from $w(\alpha,\beta,\gamma)$ and $\widehat{w}(\alpha,\beta,\gamma)$ in $R^\mrm{c}$. It is not unreasonable to expect that $\widehat{E}_{R^\mrm{c}}$ will be small when $E_{R^\mrm{c}}$ is small, however. This is because a small $E_{R^\mrm{c}}$ implies small elements of $a_1'$ associated with RGSFs having larger concentrations in $R^\mrm{c}$ and with a sufficient number of measurements $\widehat{a}'$ will be close to $a_1'$ in the sense of the $l_2$ distance (see proof of Theorem \ref{thm:restricted_so3_accuracy}).
\end{remark}

\begin{remark}
    The above result gives a theoretical foundation for CS using measurements on $R$ and its ideas can be used together with approaches tailored to give deterministic sampling patterns~\cite{culotta-lopez_compressed_2018,bangun_sensing_2020}. For example, the RGSF basis can be used and fixed measurement positions on $R$ can be iteratively added to the measurement matrix as to minimize its coherence.  
\end{remark}

\begin{remark}
    As stated, \Cref{thm:restricted_so3_accuracy} dictates the use of random polarization angle measurements when used for EM antennas. However, it is common to only use to polarization angle measurements separated by $\pi/2$~\cite{hansen_spherical_1988} when a $\mu=\pm1$ probe is used. If one desires only those two measurement positions, \Cref{thm:restricted_so3_accuracy} can be altered to work in this case. However, the BOS becomes mixed continuous (azimuth and polar angle) and discrete (polarization angle). This mixed BOS makes the discussion rather technical but does not change the fundamental nature of the result.
\end{remark}

\subsection{Sparsity in Wigner $D$-Functions and RGSFs}\label{sec:transformation_of_sparsity}
The sparsity of $s$ of the RGSF coefficients $a'$ plays an important role in Theorem \ref{thm:restricted_so3_accuracy}. As discussed earlier, we know that the Wigner $D$-function coefficients $a$ are typically sparse, using only a fraction of the $S^{m\mu}$ subspaces. The transformation $a=\sqrt{\Lambda}Ua'$ used in \eqref{eq:rotated_matrix_problem} will clearly affect the level of sparsity of $a'$ relative to $a$. Roughly speaking, the change in sparsity is because the transformation can map one coefficient of $a$ in an $S^{m\mu}$ subspace to many in the same subspace in $a'$. This effect on sparsity in the RGSF basis is straightforwardly addressed in two cases that we present in the lemmas below. The first lemma relates the RGSF sparsity to an upper bound on both $|m|$ and $|\mu|$ for which the Wigner $D$-function coefficients, $a$, are nonzero. The second lemma relates the RGSF sparsity to the sparsity in the Wigner $D$-function basis without restrictions on $m$ or $\mu$. Importantly, these lemmas show the RGSF coefficients, $a'$, will be sparse when $a$ is sparse.  Furthermore, the lemmas quantify how the sparsity level of $a'$ is related to the sparsity level of $a$.  

\begin{lemma}[Sparsity of $a'$ with Bounded $m$ and $\mu$] \label{lem:sparsity_bounded_mu_m}
    Let $a$ be the coefficient vector for $w(\alpha, \beta, \gamma)$ in the band-limited Wigner $D$-function basis with band-limit $n_{\max}$. If the Wigner $D$-function coefficients $\acoef{n}{m\mu}$ in $a$ are nonzero only for $|m|,\; |\mu| \leq m_{\max}$, then the coefficient vector $a'=\sqrt{\Lambda} U a$ in the RGSF basis is $s$-sparse with $s$ satisfying
    \begin{equation}
        s \leq N_{m_{\max}},
    \end{equation}
    where 
    \begin{equation}\label{eq:num_bandlimited_RGSF_limit_mu_m}
    \begin{split}
        N_{m_{\max}} = &\;(m_{\max}+1)(2m_{\max}+1)(2m_{\max}+3)/3 \\
        & + (n_{\max}-m_{\max})(2m_{\max}+1)^2.
        \end{split}
    \end{equation}
\end{lemma}

\tit{Proof}: The proof is given in Section VIII-B.

\begin{lemma}[Sparsity of $a'$ Given $a$ is $k$-sparse] \label{lem:sparsity_from_s_sparse}
    Let $a$ be the coefficient vector for $w(\alpha, \beta, \gamma)$ in the band-limited Wigner $D$-function basis with band-limit $n_{\max}$. If $a$ is $k$-sparse with $k\in\{1,2,\cdots,(2n_{\max}+1)^2-1\}$, then the coefficient vector $a'=\sqrt{\Lambda} U a$ in the RGSF basis is $s$-sparse with $s$ satisfying
    \begin{equation}
        s \leq N_{m_{\max}} + (k-(2m_{\max}+1)^2)(n_{\max}-m_{\max}),
    \end{equation}
    with 
    \begin{equation}
        m_{\max} = \begin{cases}
            \frac{\lfloor \sqrt{k} \rfloor-1}{2}, \; \lfloor \sqrt{k} \rfloor \text{ is odd}\\
            \frac{\lfloor \sqrt{k} \rfloor - 2}{2}, \; \lfloor \sqrt{k} \rfloor \text{ is even},
        \end{cases}
    \end{equation}
    where $N_{m_{\max}}$ is given by \eqref{eq:num_bandlimited_RGSF_limit_mu_m} and $\lfloor \cdot \rfloor$ is the floor function.
\end{lemma}
\tit{Proof}: The proof is given in Section VIII-C.

\begin{remark}
    In the lemmas above, the upper bounds on the RGSF sparsities grow with $m_{\max}^3$ and $n_{\max}m_{\max}^2$ as the highest order terms. This scaling is due to the dimension of the Wigner $D$-function basis (and RGSF basis) scaling with the $m,\;\mu$ band-limit and $n_{\max}$ as $m_{\max}^2 n_{\max}$. Since the transforming to the RGSF basis mixes coefficients within $m,\; \mu$ subspaces, the worst case sparsity is one non-zero Wigner $D$-function coefficient per $m,\;\mu$ subspace turning into full RGSF subspaces. In contrast, the best case is when all Wigner $D$-function coefficients are in a single $m,\;\mu$ subspace. This gives, at most, one full $m,\;\mu$ subspace for the RGSF coefficients.
\end{remark}

\section{Numerical Results}\label{sec:numerics}
We present numerical examples of the compressive sensing theory developed above. We begin with examples of the RGSFs and how transforming from the Wigner $D$-function basis to the RGSF basis affects sparsity. Next, we study how CS using the BOS of RGSFs is affected by the parameters of \Cref{thm:restricted_so3_accuracy} via its CS recovery phase diagram. Then, using an example field, we present how CS reconstruction according to Theorem \ref{thm:restricted_so3_accuracy} performs compared to other CS and non-CS methods in two cases. The first case is with the domain $R$ chosen to be nearly all of $\sot$ and the second is when the measurements are restricted to half of $\sot$. For these comparisons, we look at both the near-field and far-field reconstructions acquired from different CS and classical methods. Note that the far-field reconstruction is acquired by solving the coefficient vector $a$ or $a'$ for the spherical wave coefficients, which are used to calculate the field at a large distance. Next, we analyze the performance of these methods in regard to spherical wave coefficients when $R$ is restricted to half of $\mrm{SO}(3)$. Finally, we cover how the tuning of $\lambda_c$ or the addition of measurement noise affects CS reconstruction in the truncated RGSF basis.  

For the sake of simplicity, the function on $\mrm{SO}(3)$ that we reconstruct in this section is chosen to be an axisymmetric acoustic scalar field $F(r, \theta, \phi)$ (field coefficients are given in the supplementary material Section IX), for which we simulate measurements by an ideal axisymmetric probe. Here, $(r, \theta, \phi)$ are positions in spherical coordinates centered on the device. This means $w(\alpha,\beta,\gamma)$ is the rotated and translated form $F(r, \theta, \phi)$ as measured by the ideal probe. The near field measurement position for this field with wavelength $\lambda_\mrm{rad}$ is found to be $r=7 \lambda_\mrm{rad}$ and the far-field is taken at $r=2000\lambda_\mrm{rad}$. This setup is analogous to the full EM antenna case. The main differences between the acoustic and EM cases are twofold. First, the measurements $w(\alpha_j, \beta_j, \gamma_j)$ in acoustics are independent of the polarization angle. This is not so for the EM case. Second, monopole terms do not exist in the EM case, so the coefficients and explicit sensitivities of an ideal probe are different. These two differences amount to increases in the number of non-zero RGSF coefficients, but the form of the inverse problem remains otherwise unchanged. As we will see, the figures below show that when measurements are only available on $R$, CS with the truncated RGSF basis performs the best.

As a basis of comparison for our method {\bf (CS using the truncated RGSF series)}, we include recovery results from the following methods. Note, none of the following CS methods have theoretical guarantees of the form of~\Cref{thm:restricted_so3_accuracy} when using measurements only on $R$.
\begin{enumerate}
    \item \tbf{WD CS} -- CS in the Wigner $D$-function BOS using measurements on all of $\mrm{SO}(3)$~\cite{bangun_sensing_2020}
    \item \tbf{Padded FFT} -- Zero-padded FFT according to \cite{wittmann_probe-corrected_1992}
    \item \tbf{Dropped WD CS} --  CS in the Wigner $D$-function BOS using only measurements on $R$
    \item \tbf{Padded WD CS} -- CS in the Wigner $D$-function BOS using measurements on $R$ with zeros inserted for any measurement in $R^\mrm{c}$
\end{enumerate}

The WD CS method is simply for comparison to CS when full $\mrm{SO}(3)$ measurements are available. The other three methods are alternative approaches to CS when full $\mrm{SO}(3)$ measurements are not available. The Padded FFT method is a particularly important baseline for comparison; it can be considered as the classical method to maintain accuracy when measuring only a subset of $\mrm{SO}(3)$~\cite{hansen_spherical_1988}. The accuracy one should expect when using the Padded FFT method is well-known from the full EM case~\cite{hansen_spherical_1988}. In short, if measurements are only available for $R$, then the far-field reconstruction of an antenna's field is only reliable on a subset of $R$ based on the geometry of the antenna and the radius at which measurements are taken, see \cite[pp.~232--234]{hansen_spherical_1988}. 

\subsection{Parameter Selections}
In all following examples, we select $n_{\max}=20$ and $\Theta_1 = 0$ unless otherwise specified. To compute the integrals in \eqref{eq:das_rgsf_matrix} we use Gauss--Legendre quadrature. For a given $n_{\max}$, this method is numerically exact, since the product of Wigner $d$-functions is a polynomial in $\cos\beta$. Thus, it follows that a product of Wigner $d$-functions becomes a polynomial on the interval $[-1,1]$. To calculate the Wigner $d$-functions we use the recursive algorithm developed in \cite{gumerov_recursive_2015}. Additionally, eigendecompositions are computed using MATLAB\footnote{Mention of this product is not an endorsement but only serves to clarify what was done in this work.} function \texttt{eig()}. For the CS methods we use BP (for WD CS, Drop WD CS, Pad. WD CS) and QCBP (for RGSF CS) from the SPGL1 library\footnotemark[2]~\cite{berg_spgl1_2019, berg_probing_2008} with 300 non-zero simulated measurements. For the QCBP, we use the quadratic constraint $\| P w - P \Phi_1' a_1'\|\leq 0.05 \| P \eta'\|$, where $\eta'$ is known from the actual coefficients $a_2'$ and the given noise level $\eta$ (which can be zero). Lastly, for the Padded FFT method, we use $861$ simulated measurements (the minimum according to the Nyquist sampling theorem), $451$ of which are nonzero, with the remainder being padded zeros.

We note that the run time for problem setup and a single CS reconstruction using RGSFs with $n_{\max}=20$ is on the order of 30 seconds or less when done on a laptop utilizing 16 GB of RAM and one 2.3 GHz core (Intel i7)\footnotemark[2]. This run time includes the computation of the RGSF expansion coefficients in the Wigner $D$-function basis (from pre-computed integrals over $R$), which can be reduced by pre-computing the expansion coefficients.

\subsection{RGSF Examples}
\begin{figure}[t]
    \centering
    \includegraphics[width = \linewidth]{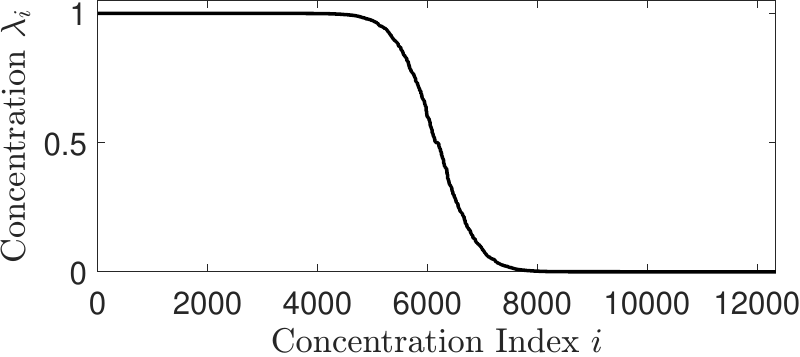} \caption{\tit{Energy Concentration of the RGSFs}. Distribution of all RGSF eigenvalues with band-limit $n_{\max}=20$.}\label{fig:rgsf_spectra}
\end{figure}

\begin{figure}[t]
    \centering
    \includegraphics[width = \linewidth]{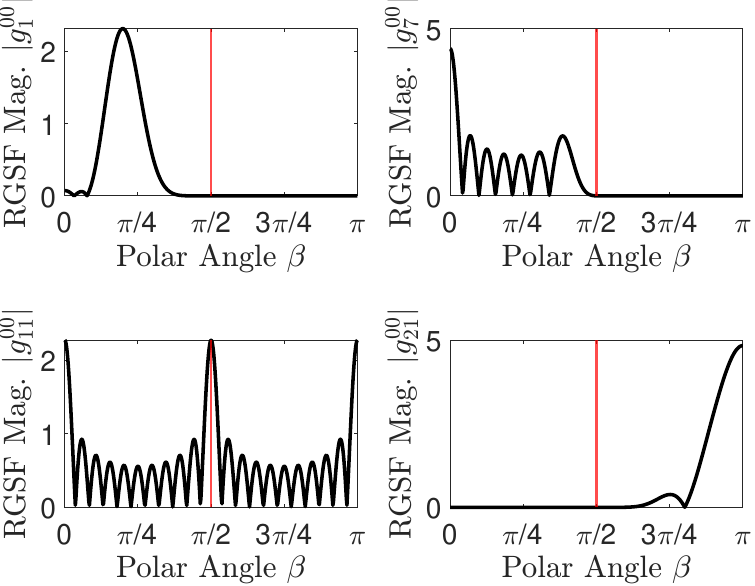} \caption{\tit{RGSF Examples}. The $\beta$ dependence of RGSF magnitudes in $S^{00}$: (a) $\t{g}{_1^{00}}$, (b) $\t{g}{_7^{00}}$, (c) $\t{g}{_{11}^{00}}$, and (d) $\t{g}{_{21}^{00}}$.}\label{fig:rgsf_plot_examples}
\end{figure}

We give examples of the sorted RGSF spectrum and the $\beta$ distribution of a few RGSFs in \Cref{fig:rgsf_spectra,fig:rgsf_plot_examples}, respectively, with $\Theta_2 = \pi/2$ and band-limit $n_{\max}=20$. Note that we have dropped the subspace indices and ordered the RGSFs according to their concentration in $R$. The spectrum shows the critical behavior that is expected for Slepian functions, where a portion of the eigenvalues are near unity and there is a rapid transition to near zero. In Figure \ref{fig:rgsf_plot_examples} we plot the magnitude of a selection of RGSFs in $S^{00}$ as a function of $\beta$. We select the first, seventh, eleventh, and last RGSFs in this subspace in order to demonstrate different levels of concentration in $R$. The first and seventh RGSFs in the subspace are almost entirely supported on $R$, the eleventh is supported on all of $\mrm{SO}(3)$, and the last one is almost entirely supported on $R^\mrm{c}$.

\subsection{Changes in Sparsity When Using RGSFs}
\begin{figure}[t]
    \centering
    \includegraphics[width = \linewidth]{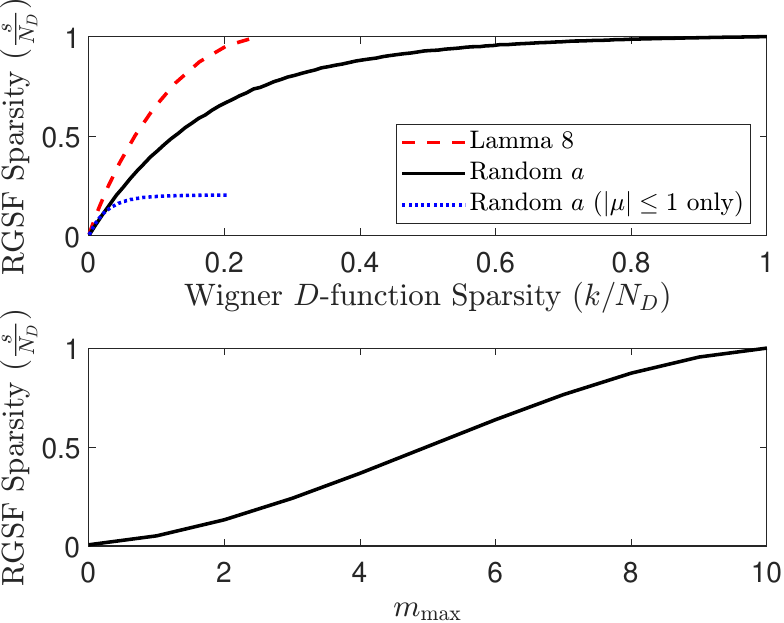} \caption{\tit{Change In Sparsity from Wigner $D$-functions to RGSFs}. Top pane: RGSF sparsity $s$ resulting from Lemma \ref{lem:sparsity_from_s_sparse}, random $k$-sparse Wigner $D$-function coefficients, and random $k$-sparse Wigner $D$-function coefficients only in $|\mu|\leq 1$ subspaces. Bottom Panel: RGSF sparsity given Wigner $D$-function coefficients satisfying Lemma \ref{lem:sparsity_bounded_mu_m}.}\label{fig:wd_to_rgsf_sparsity}
\end{figure}

As discussed above in \cref{sec:transformation_of_sparsity}, transforming from the Wigner $D$-function basis to the RGSF basis increases the sparsity level of the spherical near-field measurement inverse problem in a bounded way. In \Cref{fig:wd_to_rgsf_sparsity} we depict this change as specified by Lemma \ref{lem:sparsity_bounded_mu_m} in the bottom pane and Lemma \ref{lem:sparsity_from_s_sparse} in the top pane (dashed curve). Note, for this experiment we set $n_{\max}=10$. Lemma \ref{lem:sparsity_from_s_sparse} is the \emph{worst case} change in sparsity level, so we also depict the change in sparsity when the Wigner $D$-function coefficients, $a$, are randomly selected on all Wigner $D$-functions (top pane solid curve) or only on Wigner $D$-functions with $|\mu|\leq 1$ (dotted curve). In both cases, the values of the particular coefficients are i.i.d. random variables of the form $x+\mi y$ with both $x,y\sim \mathcal{N}(0,1)$ where $\mathcal{N}(0,1)$ the standard normal distribution and the output RGSF sparsity is averaged over 100 trials for each Wigner $D$-function sparsity. Note, the most physically relevant case is when $a$ is nonzero only for $|\mu|\leq1$, as measurements are usually conducted with probes only sensitive to low-order SWs. As can be seen in \Cref{fig:wd_to_rgsf_sparsity}, both random coefficient selections have considerably lower sparsity levels than Lemma \ref{lem:sparsity_from_s_sparse} guarantees, with $|\mu|\leq1$ being well bounded. Moreover, the results in the bottom pane show that a secondary band limit like $m_{\max}$ assumed in Lemma \ref{lem:sparsity_bounded_mu_m} results in the RGSF sparsity being well controlled unless $m_{\max}\approx n_{\max}$.

\subsection{RGSF CS Phase Diagram}
\begin{figure}[t]
    \centering
\begin{subfigure}{\linewidth}
    \centering
    \includegraphics[width = \linewidth]{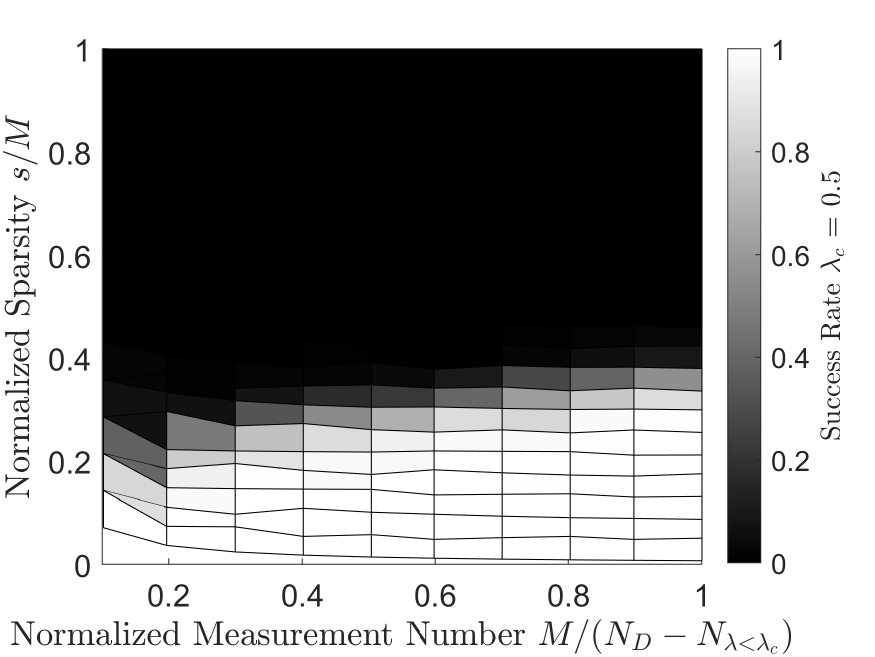}
    \caption{}
\end{subfigure}
\begin{subfigure}{\linewidth}
    \centering
    \includegraphics[width = \linewidth]{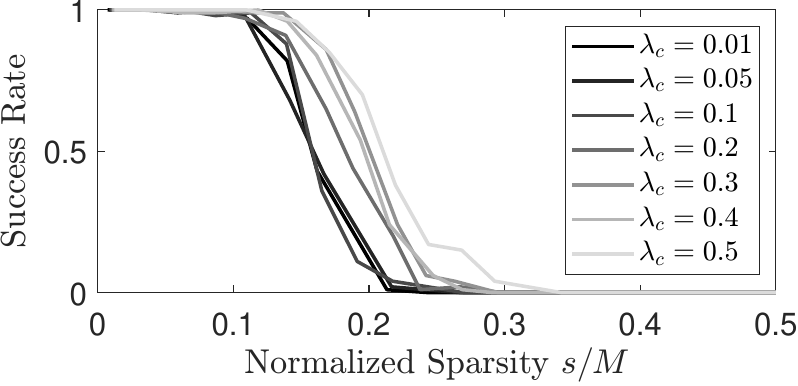}
    \caption{}
\end{subfigure}
    \caption{\tit{Phase Transition Diagrams for CS with RGSFs}. (a) Phase diagram for RGSF CS using BP to solve \eqref{eq:preconditioned_rgsf_inverse_problem} with $\eta'=0$. (b) Dependence of the phase transition on the RGSF cutoff, $\lambda_c$, for a fixed normalized measurement number of 0.6. Each plot uses $n_{\max} = 5$ and is conducted with 100 trials at each normalized sparsity and measurement number.}\label{fig:phase_diagram}
\end{figure}

\Cref{thm:restricted_so3_accuracy} gives a requirement on the number of measurements needed for CS using RGSFs to be successful. However, as is common with similar RIP-based guarantees for BOS, there are various factors affecting the number of measurements needed. To investigate how these affect recovery in practice, \Cref{fig:phase_diagram} shows the phase transition diagram generated by using BP to solve the inverse problem in \eqref{eq:preconditioned_rgsf_inverse_problem} with $\eta' = 0$ (since this encapsulates noise and modeling errors). In \Cref{fig:phase_diagram}, (a) gives the general phase diagram with $\lambda_c=0.5$ and (b) gives the distribution of the success rate, $\frac{\# \textnormal{successes}}{\# \textnormal{trials}}$, for fixed normalized sampling, $M/(N_D - N_{\lambda<\lambda_c})$, but varying $\lambda_c$. Here the solution to BP, $\widehat{a}_1$, is considered a success if $\|\widehat{a}_1' - a_1'\|/ \|a_1'\| \leq 0.001$. For \Cref{fig:phase_diagram}a-c we set $n_{\max} =5$ to allow for reasonable computation times.

In \Cref{fig:phase_diagram}a, the phase diagram is constructed by conducting 100 trials for each combination of normalized sparsity level ($s/M$) and normalized measurement number. In these trials, the support of $a_1'$ is selected uniformly at random and the coefficients are i.i.d. random variables of the form $x + \mi y$, where both $x$ and $y$ are independent and distributed according to the standard normal distribution. As can be seen in \Cref{fig:phase_diagram}a, there is a sharp transition from a success rate of near unity (i.e., uniform recovery) to a success rate of near 0 as is typically seen in CS phase diagrams like those in~\cite{rauhut_sparse_2012, bangun_sensing_2020}.

An important part of the RGSF phase diagram to understand is its dependence on the RGSF cutoff $\lambda_c$. This is because the number of measurements needed for successful CS depends on $\lambda_c^{-1}$, and ideally, the cutoff is set as low as possible. Note the presence of this parameter in \eqref{eq:preconditioned_rgsf_inverse_problem} is due to the uniform bound on the RGSFs, which is dominated by the RGSFs with small concentrations. \Cref{fig:phase_diagram}b shows the dependence of the phase transition on $\lambda_c$. For \Cref{fig:phase_diagram}b, the normalized measurement number is fixed at 0.6 and the coefficients $a_1'$ are set to have their support on the $s$ RGSFs with the smallest concentrations $\t{\lambda}{_i^{\mu m}}$ with random values of the same form as described for \Cref{fig:phase_diagram}a. The selection of this particular support is to ensure that the RGSFs that contribute the most to the selection of $\lambda_c$ are active in the CS problem. As can be seen in \Cref{fig:phase_diagram}b,  decreasing $\lambda_c$ indeed reduces the sparsity level at which the transition to successful CS recovery occurs.

\subsection{Compressive Sensing Recovery for an Example Field}
\begin{figure}[t]
\centering
\begin{subfigure}{\linewidth}
    \centering
    \includegraphics[width = \linewidth]{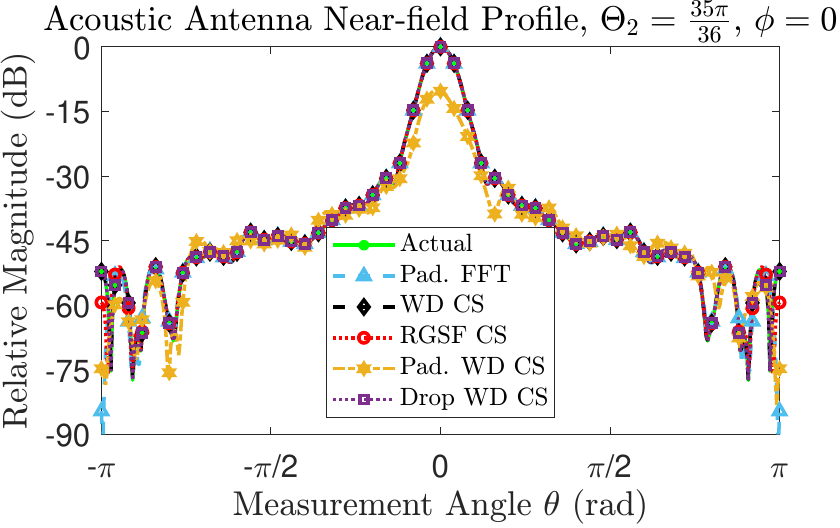}
    \caption{}
\end{subfigure}
\begin{subfigure}{\linewidth}
    \centering
    \includegraphics[width = \linewidth]{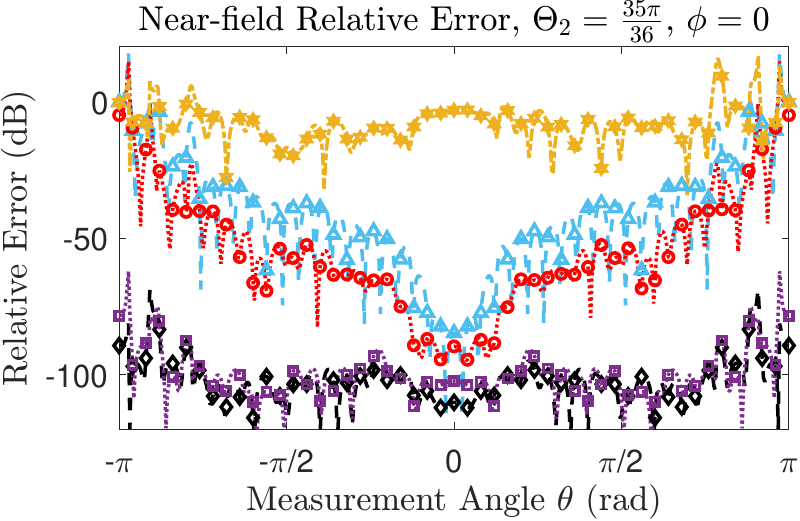}
    \caption{}
\end{subfigure}
    \caption{\tit{Near-field Reconstruction $\Theta_2=35\pi/36$}. Near-field reconstruction (a) and relative error (b) when $R$ is nearly all of $\mrm{SO}(3)$. (a) and (b) share the legend in (a).}\label{fig:nf_35pi36}
\end{figure}

\begin{figure}[t]
\centering
\begin{subfigure}{\linewidth}
    \centering
    \includegraphics[width = \linewidth]{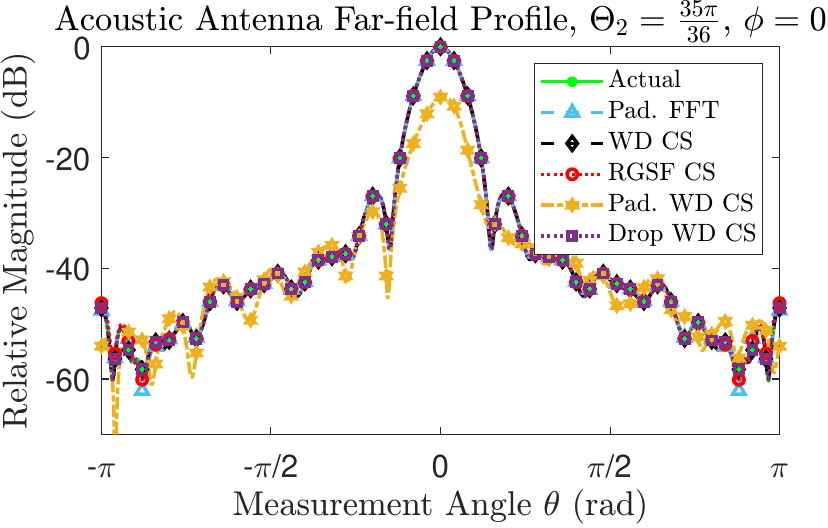}
    \caption{}
\end{subfigure}
\begin{subfigure}{\linewidth}
    \centering
    \includegraphics[width = \linewidth]{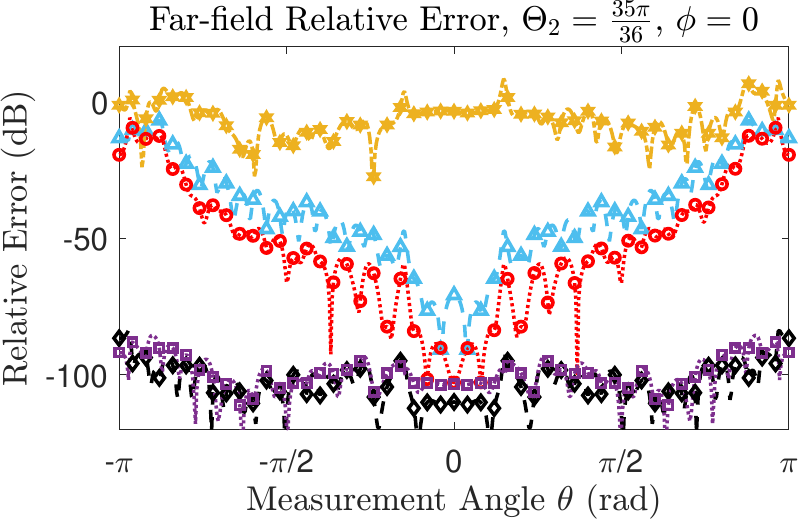}
    \caption{}
\end{subfigure}
    \caption{\tit{Far-field Reconstruction $\Theta_2=35\pi/36$}. Far-field reconstruction (a) and relative error (b) when $R$ is nearly all of $\mrm{SO}(3)$. (a) and (b) share the legend in (a).}\label{fig:ff_35pi36}
\end{figure}

To verify that CS recovery using RGSFs becomes accurate for the function on the entirety of $\mrm{SO}(3)$ as $R$ becomes close to all of $\mrm{SO}(3)$, we investigate CS recovery using the RGSFs with $\Theta_2=35\pi/36$, or a loss of 5 degrees of measurement availability in $\beta$. We select $\lambda_c=0.5$ as to remove at least several RGSFs from the reconstructions. We must select a moderately sized $\lambda_c$ because most of the RGSFs will have at least some energy in $R$ due to the small size of $R^\mrm{c}$ and a relatively small band-limit. In \Cref{fig:nf_35pi36,fig:ff_35pi36}, we plot the near-field and far-field reconstructions and relative error for all methods. Note, the relative error for WD CS is not numerically zero since there is a finite error tolerance using BP in SPGL1. The reconstructions are plotted in dB in terms of magnitude relative to the actual field for a given $\phi$ slice, i.e.,
\begin{equation}
    \text{Relative Magnitude} = 20\log_{10}\left(\frac{\left|\widehat{F}\right|}{\max_{\theta}|F|}\right),
\end{equation}
and the relative error of the field in dB is given by
\begin{equation}
    \text{Relative Error} = 20\log_{10}\left(\frac{|F-\widehat{F}|}{|F|}\right),
\end{equation}
where $\widehat{F}$ is the estimated reconstruction of $F$. From $\theta = -\pi/2$ to $\theta = \pi/2$, the near-field reconstructions are accurate to around 50 dB for all methods except the Padded WD CS method. However, for the RGSF CS and Padded FFT methods, we see a decline in accuracy beyond this range; the relative error increases to near 0 dB in $R^{\mrm{c}}$. This is expected since the RGSF reconstruction lacks several functions concentrated on $R^\mrm{c}$ and the padded FFT has zeros inserted in this region. The far-field reconstructions are similar to those of the near field.

\begin{figure}[t]
\centering
\begin{subfigure}{\linewidth}
    \centering
    \includegraphics[width = \linewidth]{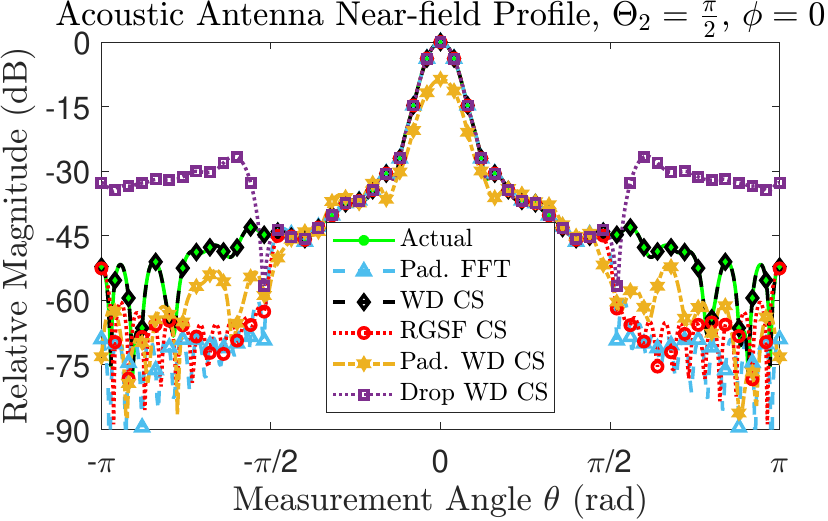}
    \caption{}
\end{subfigure}
\begin{subfigure}{\linewidth}
    \centering
    \includegraphics[width = \linewidth]{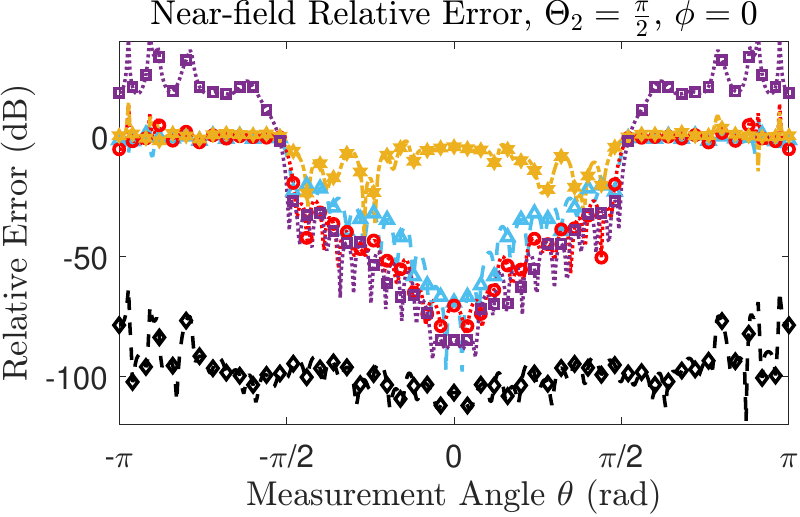}
    \caption{}
\end{subfigure}
    \caption{\tit{Near-field Reconstruction $\Theta_2=\pi/2$}. Near-field reconstruction (a) and relative error (b) when $R$ is half of $\mrm{SO}(3)$. (a) and (b) share the legend in (a).}\label{fig:nf_pi2}
\end{figure}

\begin{figure}[t]
\centering
\begin{subfigure}{\linewidth}
    \centering
    \includegraphics[width = \linewidth]{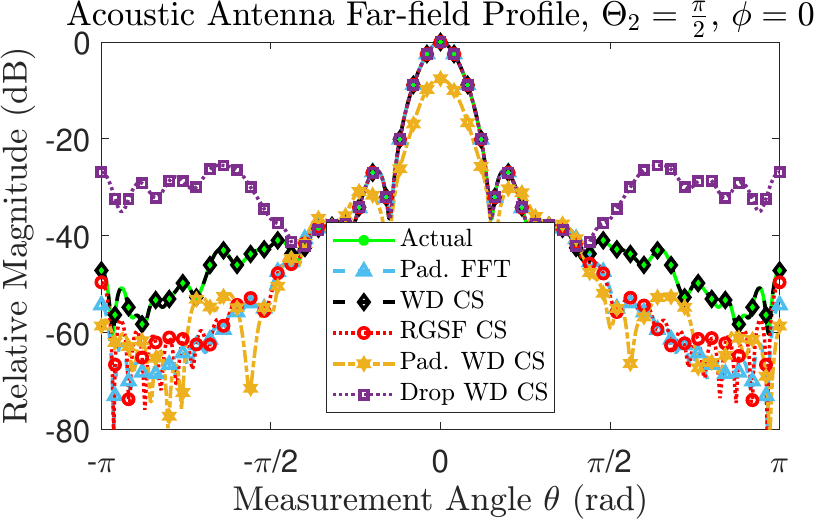}
    \caption{}
\end{subfigure}
\begin{subfigure}{\linewidth}
    \centering
    \includegraphics[width = \linewidth]{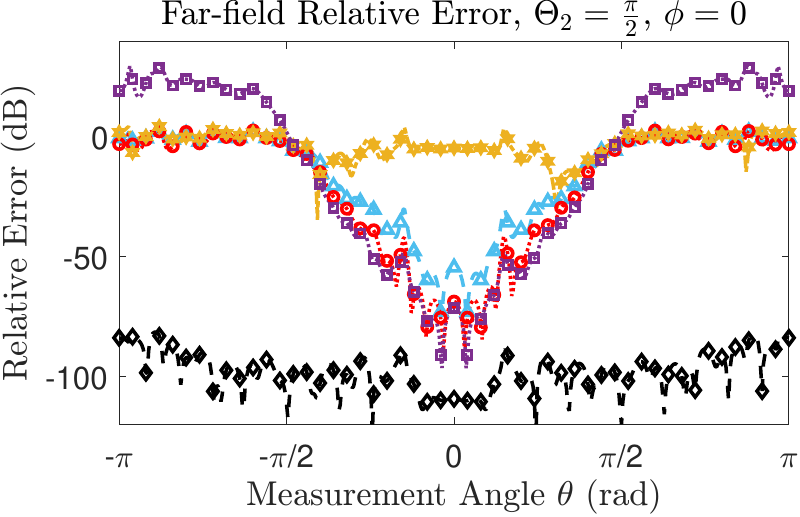}
    \caption{}
\end{subfigure}
    \caption{\tit{Far-field Reconstruction $\Theta_2=\pi/2$}. Far-field reconstruction (a) and relative error (b) when $R$ is half of $\mrm{SO}(3)$. (a) and (b) share the legend in (a).}\label{fig:ff_pi2}
\end{figure}

We now compare the reconstruction methods when a much larger portion of $\mrm{SO}(3)$ is inaccessible. In particular, we select $\Theta_2=\pi/2$ and $\lambda_c = 0.05$. The near-field and far-field reconstructions are shown in \Cref{fig:nf_pi2,fig:ff_pi2}, respectively. Recall that with such a restriction on $R$, we only expect accurate field reconstructions within $R$. For this case of $R$ and $\lambda_c$, we see that the near-field reconstructions for all but the Padded WD CS methods are accurate in nearly all of $R$, with deterioration near the edges of $R$. The far field shows a similar behavior, with the degradation occurring slightly more rapidly. In $R^\mrm{c}$ we see that the Dropped WD CS method overestimates the field by a large margin while the other methods have field values near zero as expected.

\begin{figure}[t]
    \centering
    \includegraphics[width = \linewidth]{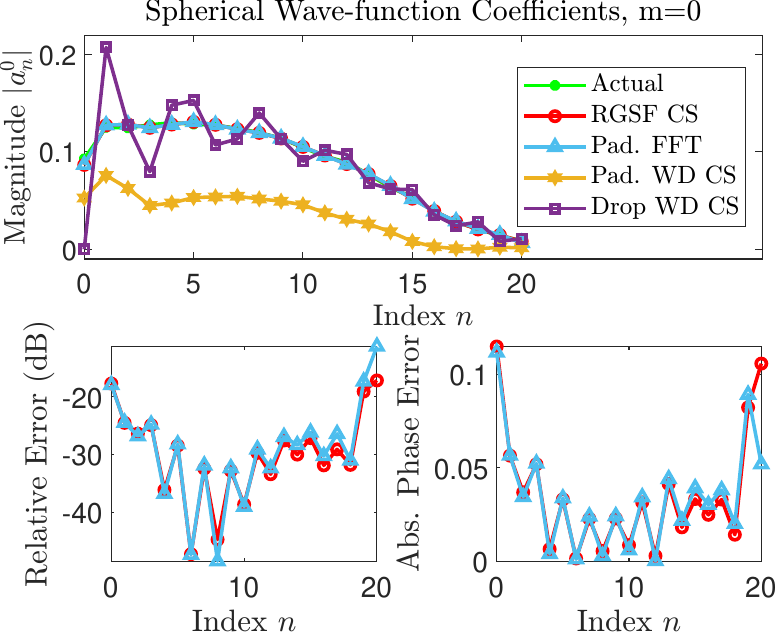}
    \caption{\tit{Reconstructed SW Coefficients $\Theta_2=\pi/2$}. (Top) Reconstructed SW wave coefficients $a_n^m$ with $m=0$. (Bottom left) SW coefficient relative error for RGSF CS and Padded FFT reconstructed SW coefficients with $m=0$. (Bottom right) Absolute Phase error in radians for RGSF CS and Padded FFT reconstructed SW coefficients with $m=0$. All figures share the same legend.} \label{fig:pwc_plots}
\end{figure}

To investigate the performance of each restricted measurement method further, we plot the $m=0$ spherical wave coefficients $a_n^m$ in \Cref{fig:pwc_plots}. Since the field is axisymmetric, the $a_n^m$ should be nonzero only when $m = 0$. The Padded FFT method gives numerical zeros for all SW coefficients with $m \neq 0$. On the other hand, the other methods have nonzero coefficients with $m \neq 0$. For the RGSF CS case, the energy contained in these coefficients is less than $3\times10^{-4}$ of the total energy, for Dropped WD CS it is less than $4\times10^{-5}$, and for Padded WD CS it is $3\times10^{-2}$ of the total energy. Figure \ref{fig:pwc_plots} demonstrates that the RGSF CS and Padded FFT methods have good performance for both SW coefficient relative error and absolute phase error in radians. However, the Dropped WD CS and Padded WD CS methods have large errors visible just by looking at the coefficient magnitudes (\Cref{fig:pwc_plots}a). It is interesting to note that the Dropped WD CS method appears to oscillate around the actual coefficients. From \Cref{fig:nf_pi2,fig:ff_pi2,fig:pwc_plots} we see that the RGSF CS method performs the best compared to the other restricted measurement methods and the Padded FFT method as a close second.

\begin{figure}[t]
\centering
\begin{subfigure}{0.45\textwidth}
    \centering
    \includegraphics[width = \linewidth]{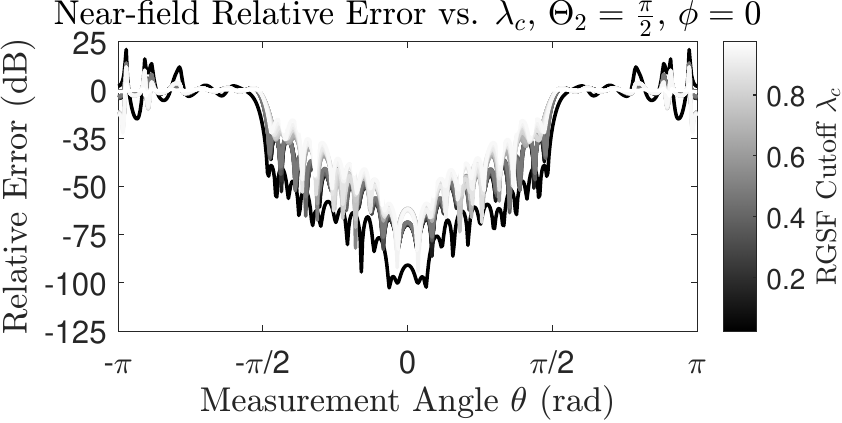}
    \caption{}
\end{subfigure}
\begin{subfigure}{0.45\textwidth}
    \centering
    \includegraphics[width = \linewidth]{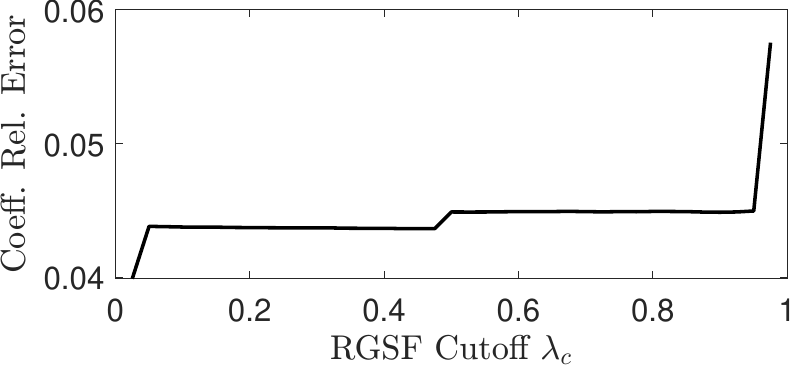}
    \caption{}
\end{subfigure}
    \caption{\tit{RGSF Recovery vs.\ $\lambda_c$}. (a) The relative error for the RGSF CS near-field reconstruction and (b) Wigner $D$-function coefficient relative error. In (a) the darker curves indicate smaller values of $\lambda_c$.}\label{fig:error_vs_lambda_c}
\end{figure}

We now analyze the performance of the RGSF CS method as we vary $\lambda_c$. For this study, we use $\Theta_2=\pi/2$ and vary $\lambda_c$ from 0.05 to 0.95 in increments of 0.025. \Cref{fig:error_vs_lambda_c}a plots the near-field relative error and \Cref{fig:error_vs_lambda_c}b plots the relative error between the actual Wigner $D$-function coefficients $a$ and those from RGSF reconstruction,
\begin{equation}\label{eq:norm_coeff_err}
    \mathrm{Coeff.\; Relative \; Error} = \frac{\|a-\widehat{a}\|}{\|a\|},
\end{equation}
where $\widehat{a} = U^* \sqrt{\Lambda}^{-1} \widehat{a}'$, and $\widehat{a}'$ is constructed from the recovered $\widehat{a}_1'$ and $\widehat{a}_2'=0$. In \Cref{fig:error_vs_lambda_c}a, the lighter the curve corresponds to the larger cutoff $\lambda_c$. Thus, as $\lambda_c$ decreases, the reconstruction performance improves. Similarly, if we look at the relative error of the field and $\widehat{a}$, the lower $\lambda_c$ values provide better reconstruction performance in the RGSF basis. Not depicted here, however, is the case where $\lambda_c$ becomes very small. Experiments have shown that if $\lambda_c$ is taken to be too small, the reconstruction begins to grow in $R^\mrm{c}$ to values well above the true field. This is likely due to numerical accuracy problems in the basis pursuit method; functions with small $\t{\lambda}{_i^{\mu m}}$ contribute so little to the measurements in $R$ that minor fluctuations in their coefficients can lead to possibly sizable changes in the field on $R^\mrm{c}$. In summary, setting $\lambda_c$ too large  can lead to poor reconstructions depending on the amount of energy the field has in $R^\mrm{c}$ while selecting $\lambda_c$ too small leads to errors due to an effective freedom to add RGSFs with small concentrations without a large effect on the measured portions of the field. Thus, two factors should go into picking an ideal $\lambda_c$, an idea of how well concentrated the field is (the larger this is, the larger $\lambda_c$ can be) and the maximum magnitude of the RGSFs discarded on $R$. 

\begin{figure}[t]
    \centering
    \includegraphics[width = \linewidth]{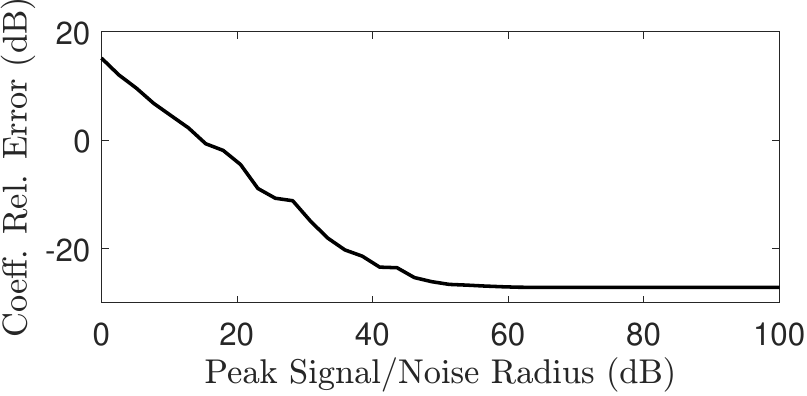} \caption{\tit{Wigner $D$-function Coefficient Error vs. Noise Level.} Relative error as a function of measurement noise level for the Wigner $D$-function coefficients of the example field reconstructed using RGSF CS with $\Theta_2=\pi/2$ and $\lambda_c = 0.05$.}\label{fig:noisy_recons}
\end{figure}

We last analyze the performance of the RGSF CS method with respect to additive measurement noise. \Cref{thm:restricted_so3_accuracy}, in particular \eqref{eq:D_coef_error}, states that the coefficient relative error of the reconstruction $\widehat{a}$ increases with increasing noise. In \Cref{fig:noisy_recons} we plot the normalized coefficient error, \eqref{eq:norm_coeff_err}, as we increase the magnitude of the additive measurement noise. Again we have selected $\Theta_2=\pi/2$ and $\lambda_c = 0.05$. Each entry of the measurement noise vector is i.i.d. and uniformly distributed on the complex unit disc with radius $r_{\mrm{noise}}$ that gives the specified Peak Signal to Noise Radius ratio in dB,
\begin{equation}
    \frac{\mathrm{Peak\; Signal}}{\mrm{\; Noise \;Radius}} = 20\log_{10}\left( \frac{\max_{\alpha,\beta,\gamma} w(\alpha,\beta,\gamma)}{r_{\mrm{noise}}} \right).
\end{equation}
As can be seen in \Cref{fig:noisy_recons}, if the noise level increases (Peak Signal/Noise Radius  decreases), the normalized coefficient error increases. Note, noise levels below 40 dB are such that the noise would have a magnitude approaching the size of the barely forming side-lobes in the near-field. This means these noise levels are quite large; typically one expects noise well below the side-lobes. In the more physical case where $\mathrm{Peak Signal/Noise Radius} \approx 40$ dB, RGSF CS coefficient accuracy is about on par with the noise-less case depicted in \Cref{fig:error_vs_lambda_c}.

\subsection{Remarks on Reconstruction Performance}
The results in \Cref{fig:nf_35pi36,fig:ff_35pi36} support the expectation that the ideal case for reconstruction is when all of $\mrm{SO}(3)$ is available for measurements and we can perform CS using the Wigner $D$-function basis. However, when only $R$ is available for measurement and $R^\mrm{c}$ is sizable, \Cref{fig:nf_pi2,fig:ff_pi2,fig:pwc_plots} show that CS with the truncated RGSF basis provides the best performance for both the near-field and far-field reconstructions. Specifically, the Padded Wigner $D$-function approach fails to accurately reconstruct the main beam. This is likely due to the zero padding procedure weighting points for which we do not know the field. The Dropped Wigner $D$-function approach performs well in $R$ but vastly overestimates the field in $R^\mrm{c}$, hinting that the method recovers incorrect SW coefficients (this is confirmed by \Cref{fig:pwc_plots}). This is likely due to having enough freedom in the Wigner $D$-function basis to fit the field but at the cost of an effectively arbitrary field on $R^\mrm{c}$ (similar to spline fitting with free-floating ends). The FFT and RGSF methods both perform similarly, though the RGSF method requires fewer measurements. However, close inspection shows that the near-field reconstruction from the FFT method is subject to Gibbs phenomena near the edge of $R$ that is worse than the ripples from the RGSF method. It is also worth noting that the CS method using the RGSFs is the only CS method here that guarantees accurate recovery and the fact that the Dropped Wigner $D$-function method performs well in $R$ is a fortunate outcome. Lastly, the investigations in~\Cref{fig:error_vs_lambda_c,fig:noisy_recons} show that, for the tested field, the RGSF CS method is not all too sensitive to $\lambda_c$ and the addition of noise, at least when $\lambda_c$ or the size of the noise are not too extreme.

\section{Conclusion}
Motivated by the limitations of measurement platforms in spherical near-field to far-field transformations, we have established a theoretical guarantee for the success and accuracy of compressive sensing on a restricted spherical domain for band-limited functions on $\sot$, \Cref{thm:restricted_so3_accuracy}. The guarantee uses a BOS of localized Slepian functions (RGSFs with concentration less than $\lambda_c$) on the measurement domain to limit the growth of the BOS bound and thus the number of measurements for the CS recovery is guaranteed. Numerically generated phase diagrams showed that this dependence on RGSF concentration cutoff, $\lambda_c$, is present in practice, but the degree to which it affects real-world application is only qualitatively understood. Future work can aim to further quantify and minimize the dependence measurement numbers have on $\lambda_c$ in CS guarantees like \Cref{thm:restricted_so3_accuracy}. It is worth noting that the bound for the RGSFs is over the domain $\mrm{SO}(3)$, while the bound is only needed for the sub-domain $R$. Thus, improved bounds for the RGSFs as well as Slepian functions on more general domains would be an interesting topic for future work.

The method of constructing the RGSFs in this paper is based upon performing many integrations to construct the matrix $\msc{D}$ in the functions' defining eigenvalue problem. Many domains like the sphere have operator constructions that allow for defining this eigenvalue problem without integration~\cite{simons_scalar_2015}. Future work might focus on improving the RGSF constructions by attaining an explicit form of such an operator for the RGSFs. Moreover, finding a fast Slepian transform for the RGSFs, as in \cite{karnik_fast_2019}, would also improve upon the numerical performance of the restricted measurement domain compressive sensing problem as treated here. 

In our tests, we saw that CS with the truncated RGSF basis can have relative error comparable to the classical restricted domain method with roughly $60\%$ as many measurements. Additionally, in our tests, CS with the truncated RGSF basis outperformed the tested forms of measurement restricted CS using Wigner $D$-functions. These Wigner $D$-function methods tend to give larger errors in reconstructed SW coefficients and have no theoretical guarantee of recovery, unlike the RGSF based method. Our numerical experiments also showed that CS results using the RGSFs improve as the number of Slepian functions kept for recovery increases. However, the reconstructions suffered beyond the measurement domain. Future work can look to include more \tit{a priori} information about a field's coefficient structure in the RGSF basis so that performance improves in regions beyond the measurable domain while also further decreasing the number of measurements needed for accurate reconstructions. 

Lastly, as presented, the RGSF-based CS method uses random measurement positions. It is sometimes desirable to instead have deterministic measurement positions. To that end, future research can investigate the use of RGSFs and samples on $R$ paired with now theoretical approaches, iterative coherence minimization, or other coherence-minimizing algorithms for constructing deterministic measurement matrices from known basis functions.


%

\section*{Acknowledgment}
This work was partially supported by NSF grants CCF-1704204 and CCF-2106834.

\ifCLASSOPTIONcaptionsoff
  \newpage
\fi



\bibliographystyle{IEEEtran}
\bibliography{references}

%








\ifarXiv
    \foreach \x in {1,...,\numbersupplementpages}
    {
        \clearpage
        \includepdf[pages={\x}]{\supplementfilename}
    }
\fi

\end{document}


%
\title{Supplementary Material for Compressive Sensing with Wigner $D$-functions on Subsets of the Sphere}
%
%
%

\author{Marc~Andrew~Valdez,
        Alex~J.~Yuffa,
        Michael~B.~Wakin,~\IEEEmembership{Fellow,~IEEE}%
\thanks{M.~A.~Valdez is with the Department of Electrical Engineering, Colorado School of Mines, Golden, Colorado 80401, USA and the National Institute of Standards and Technology, Boulder, Colorado 80305, USA (e-mail: mvaldez@mines.edu)}%
\thanks{A.~J.~Yuffa is with the National Institute of Standards and Technology, Boulder, Colorado 80305, USA (e-mail: alex.yuffa@nist.gov).}%
\thanks{ M.~B.~Wakin is with the Department of Electrical Engineering, Colorado School of Mines, Golden, Colorado 80401, USA (e-mail: mwakin@mines.edu).}%
}
\maketitle

\begin{abstract}
    This supplementary material contains the derivations and proofs for the main results in Compressive Sensing with Wigner $D$-functions on Subsets of the Sphere.
\end{abstract}


%
\IEEEpeerreviewmaketitle

\setcounter{theorem}{8}
\setcounter{section}{6}
\setcounter{equation}{57} 
\setcounter{figure}{8} 

\section{Derivations of Properties of the RGSFs}
\subsection{RGSF Orthogonality Relations} \label{sec:ortho_derivation}
The orthogonality relation in \eqref{eq:SO3_orthogonality} can be seen from the following calculation. Using \eqref{eq:rg_slepian_expansion} in the inner product we find
\begin{equation*}
    \inner{\t{g}{_{i}}}{\t{g}{_{i'}}}{\sot} = \sum_{n,\mu,m}^{n_{\max}} \sum_{n',\mu',m'}^{n_{\max}} \t{\widetilde{g}}{_{in}^{\mu m}} \overline{\t{\widetilde{g}}{_{i'n'}^{\mu' m'}}} \inner{\WD{n}{\mu}{m}}{\WD{n'}{\mu'}{m'}}{\sot}
\end{equation*}
By \eqref{eq:normalized_wigner_D_orthogonality} and the orthonormality of the $\mbf{g}_i$ we arrive at \eqref{eq:SO3_orthogonality}, i.e.,
\begin{align*}
    \inner{\t{g}{_{i}}}{\t{g}{_{i'}}}{\sot} &= \sum_{n,\mu,m}^{n_{\max}} \sum_{n',\mu',m'}^{n_{\max}} \t{\widetilde{g}}{_{in}^{\mu m}} \overline{\t{\widetilde{g}}{_{i'n'}^{\mu' m'}}} \t{\delta}{_{nn'}}\t{\delta}{_{mm'}}\t{\delta}{_{\mu\mu'}} \\
    &= \sum_{n,\mu,m}^{n_{\max}} \t{\widetilde{g}}{_{in}^{\mu m}} \overline{\t{\widetilde{g}}{_{i'n}^{\mu m}}} \\
    &= \t{\delta}{_{ii'}}.
\end{align*}
The orthogonality relation over the region $R$ in \eqref{eq:R_orthogonality} can be seen from a similar calculation. We begin by using \eqref{eq:rg_slepian_expansion} to obtain
\begin{align*}
    \inner{\t{g}{_{i}}}{\t{g}{_{i'}}}{R} = \sum_{n',\mu',m'}^{n_{\max}} \overline{\t{\widetilde{g}}{_{i'n'}^{\mu' m'}}} \sum_{n,\mu,m}^{n_{\max}} \t{\widetilde{g}}{_{in}^{\mu m}} \inner{\WD{n}{\mu}{m}}{\WD{n'}{\mu'}{m'}}{R}.
\end{align*}
We can recognize this as a matrix calculation using $\msc{D}$ and its eigenvectors. Substituting this in and then using the orthonormality of the eigenvectors we get \eqref{eq:R_orthogonality}, i.e.,
\begin{align*}
    \inner{\t{g}{_{i}}}{\t{g}{_{i'}}}{R} = &= \t{\mbf{g}}{_{i'}^*} \msc{D} \t{\mbf{g}}{_i} \\
    &= \lambda_i \t{\mbf{g}}{_{i'}^*} \t{\mbf{g}}{_i}\\
    &= \lambda_i \t{\delta}{_{ii'}}.
\end{align*}

\subsection{Expansion of Wigner $D$-Functions in RGSFs} \label{sec:inverse_derivation}
The inverse transformation from RGSFs to Wigner $D$-functions can be seen from the orthogonality of the eigenvectors of the $\t{\msc{D}}{^{\mu m}}$ matrix. First we write \eqref{eq:das_rgsf_expansion} in matrix form, i.e.,
\begin{equation*}
    \begin{bmatrix}
        \t{g}{_{1}^{\mu m}}(\alpha, \beta, \gamma) \\
        \t{g}{_{2}^{\mu m}}(\alpha, \beta, \gamma) \\
        \vdots \\
        \t{g}{_{i_{\max}}^{\mu m}}(\alpha, \beta, \gamma)
    \end{bmatrix} = \begin{bmatrix}
        \t{\mbf{g}}{_{1}^{\mu m}}^T\\
        \t{\mbf{g}}{_{2}^{\mu m}}^T \\
        \vdots \\
        \t{\mbf{g}}{_{i_{\max}}^{\mu m}}^T
    \end{bmatrix} \begin{bmatrix}
        \WD{n_{\min}}{\mu}{m}(\alpha, \beta, \gamma) \\
        \WD{n_{\min}+1}{\mu}{m}(\alpha, \beta, \gamma) \\
        \vdots \\
        \WD{n_{\max}}{\mu}{m}(\alpha, \beta, \gamma)
    \end{bmatrix},
\end{equation*} 
where $i_{\max} = n_{\max}-n_{\min}+1$. Next, we note that the matrix of eigenvectors $\t{\mbf{g}}{_{i}^{\mu m}}^T$ is orthogonal (by the eigendecomposition of $\msc{D}^{\mu m}$) to obtain
\begin{equation*}
    \begin{bmatrix}
        \t{\mbf{g}}{_{1}^{\mu m}} &  \cdots & \t{\mbf{g}}{_{i_{\max}}^{\mu m}}
    \end{bmatrix} \begin{bmatrix}
        \t{g}{_{1}^{\mu m}}(\alpha, \beta, \gamma) \\
        \t{g}{_{2}^{\mu m}}(\alpha, \beta, \gamma) \\
        \vdots \\
        \t{g}{_{i_{\max}}^{\mu m}}(\alpha, \beta, \gamma)
    \end{bmatrix} =  \begin{bmatrix}
        \WD{n_{\min}}{\mu}{m}(\alpha, \beta, \gamma) \\
        \WD{n_{\min}+1}{\mu}{m}(\alpha, \beta, \gamma) \\
        \vdots \\
        \WD{n_{\max}}{\mu}{m}(\alpha, \beta, \gamma)
    \end{bmatrix}.
\end{equation*} 
Rewriting this as a sum we get the desired result, namely,
\begin{equation*}
    \WD{n}{\mu}{m}(\alpha, \beta, \gamma) = \sum_{i=1}^{n_{\max}-n_{\min} +1} \t{\widetilde{g}}{_{in}^{\mu m}} \t{g}{_{i}^{\mu m}}(\alpha, \beta, \gamma).
\end{equation*}

\section{Proofs for Main Results} \label{sec:proofs}

\subsection{Proof of Proposition \ref{prop:rgsf_bound}} \label{sec:bound_prop_proof}
Using the bound on the Wigner-$D$ functions in \eqref{eq:Wigner_D_Bound} along with \eqref{eq:das_rgsf_expansion} we find 
\begin{equation*}
    \left| \sqrt{\frac{\sin \beta}{\t{\lambda}{_{i}^{\mu m}}}}\t{g}{_{i}^{\mu m}}(\alpha, \beta, \gamma)\right| \leq \frac{C' N_D^{1/12}}{\sqrt{\t{\lambda}{_{i}^{\mu m}}}} \left|\sum_{n=\max\left(|\mu|,|m|\right)}^{n_{\max}}\t{\widetilde{g}}{_{in}^{\mu m}}\right|
\end{equation*}
and from the triangle inequality we obtain
\begin{equation}
    \label{eq:after_triangle_inequality}
    \left| \sqrt{\frac{\sin \beta}{\t{\lambda}{_{i}^{\mu m}}}}\t{g}{_{i}^{\mu m}}(\alpha, \beta, \gamma)\right| \leq \frac{C' N_D^{1/12}}{\sqrt{\t{\lambda}{_{i}^{\mu m}}}}  \sum_{n=\max\left(|\mu|,|m|\right)}^{n_{\max}}\left|\t{\widetilde{g}}{_{in}^{\mu m}}\right|.
\end{equation}  
We can recognize the sum on the right-hand side of \eqref{eq:after_triangle_inequality} as $\|\t{\mbf{g}}{_i^{\mu m}}\|_1$. Noting that $\|\t{\mbf{g}}{_i^{\mu m}}\|_2=1$ (since they are orthonormal eigenvectors of the symmetric matrix $\msc{D}^{\mu m}$), we can use the special case of H\"{o}lder's inequality that given $v\in \mathbb{C}^\mcl{N}$, $\|v\|_1 \leq \sqrt{\mcl{N}}\|v\|_2$ to find 
\begin{equation}
    \left| \sqrt{\frac{\sin \beta}{\t{\lambda}{_{i}^{\mu m}}}} \t{g}{_{i}^{\mu m}}(\alpha, \beta, \gamma)\right| \leq \frac{C' N_D^{1/12}}{\sqrt{\t{\lambda}{_{i}^{\mu m}}}} \sqrt{n_{\max}+1}
\end{equation}
since the largest dimension of any $\msc{D}^{\mu m}$ is $n_{\max}+1$. In \eqref{eq:num_D_funcs}, we can see its right-hand side has two terms greater than $n_{\max}+1$. Replacing $2n_{\max}+1$ with $n_{\max}+1$ so we have an inequality version of \eqref{eq:num_D_funcs}, then replacing $2n_{\max}+3$ with $2n_{\max}+2$, factoring out the 2, and then solving for $\sqrt{n_{\max}+1}$ we find that $\sqrt{n_{\max}+1} \leq \left(\frac{3}{2}\right)^{1/6}N_D^{1/6}$. Thus, we obtain
\begin{equation}\label{eq:intermediate_rgsf_bound}
    \left| \sqrt{\frac{\sin \beta}{\t{\lambda}{_{i}^{\mu m}}}} \t{g}{_{i}^{\mu m}}(\alpha, \beta, \gamma)\right| \leq \frac{C'' N_D^{1/4}}{\sqrt{\t{\lambda}{_{i}^{\mu m}}}},
\end{equation}
with $C''=\left(\frac{3}{2}\right)^{1/6}C'$. Thus, denoting $\lambda_\text{min}$ as the minimum $\lambda_i^{\mu m}$ associated with any RGSF in a subset $\mcl{S}$ of all band-limited RGSFs on $R$ yields
\begin{equation}
    \sup_{{g}{_i^{\mu m}}\in\mcl{S}} \left\| \sqrt{\frac{\sin \beta}{\t{\lambda}{_i^{\mu m}} }} \t{g}{_i^{\mu m}}(\alpha, \beta, \gamma)\right\|_\infty \leq \frac{C'' N_D^{1/4}}{ \sqrt{ \lambda_\text{min} } } 
\end{equation}
for some constant $C''$. 

\subsection{Proof of Lemma \ref{lem:sparsity_bounded_mu_m}}\label{sec:lem_bounded_mu_m_proof}
Assume the transformation $a'=\sqrt{\Lambda}Ua$ maximally mixes the coefficients in the $S^{m\mu}$ subspace, i.e., any one element of $a$ corresponding to a function in $S^{m\mu}$ is mapped to a nonzero coefficient on all RGSFs in the subspace. This scenario corresponds to the worst-case increase in nonzero entries going from $a$ to $a'$. Thus, we need to calculate the total number of band-limited RGSFs with $\mu,\;m\leq m_{\max}$. This is the same as calculating the number of Wigner $D$-functions with the same index constraints, i.e., $\mu, m \leq m_{\max}$. There are $(m_{\max}+1)(2m_{\max}+1)(2m_{\max}+3)/3$ RGSFs for $n\leq m_{\max}$. The remaining RGSFs satisfy $n>m_{\max}$ and $m,\;\mu \leq m_{\max}$, giving $n_{\max}-m_{\max}$ possible values of $n$. For each $n$ there are $2m_{\max}+1$ functions. Thus, there are $(n_{\max}-m_{\max})(2m_{\max}+1)$ remaining functions for $m_{\max}<n\leq n_{\max}$. Adding these gives the desired result. 

\subsection{Proof of Lemma \ref{lem:sparsity_from_s_sparse}}\label{sec:lem_sparsity_from_s_sparse}
Take the same assumptions as those in \Cref{sec:lem_bounded_mu_m_proof} and further assume that each nonzero element of $a$ belongs to different $S^{m\mu}$. This is the worst-case increase in nonzero entries transforming from $a$ to $a'$. Any other case has more than one element of $a$ mapped to the same subspace, so the sparsity level of $a'$ would be less than the case described. Note that we restrict $k$ to be less than $(2n_{\max}+1)^2$ because this is the total number of $S^{m\mu}$ subspaces, and having $k$ any larger would result in a dense $a'$ since all subspaces are used. 

Now, let the nonzero elements of $a$ be indexed as $a_i$ so that $i$ increases as the associated $S^{m\mu}$ subspace dimension decreases. So $a_1$ is in the largest $S^{m\mu}$ subspace, $a_2$ is in the next largest, and so on. Thus, to get the sparsity of $a'$, $s$, we sum the dimensions of the first $k$ largest $S^{m\mu}$ subspaces. 
\begin{figure}
    \centering
    \begin{tikzpicture}
        \filldraw[red] (0,0) circle (2pt) node[anchor=west] {};
        \filldraw[black] (1,0) circle (2pt) node[anchor=west] {};
        \filldraw[black] (0,1) circle (2pt) node[anchor=west] {};
        \filldraw[black] (-1,0) circle (2pt) node[anchor=west] {};
        \filldraw[black] (0,-1) circle (2pt) node[anchor=west] {};
        \filldraw[black] (1,1) circle (2pt) node[anchor=west] {};
        \filldraw[black] (-1,1) circle (2pt) node[anchor=west] {};
        \filldraw[black] (-1,-1) circle (2pt) node[anchor=west] {};
        \filldraw[black] (1,-1) circle (2pt) node[anchor=west] {};
        \filldraw[black] (2,0) circle (2pt) node[anchor=west] {$(\mu=2,m=0)$};
        \filldraw[black] (0,2) circle (2pt) node[anchor=west] {};
        \filldraw[black] (-2,0) circle (2pt) node[anchor=west] {};
        \filldraw[black] (0,-2) circle (2pt) node[anchor=west] {};
        \filldraw[black] (2,1) circle (2pt) node[anchor=west] {};
        \filldraw[black] (1,2) circle (2pt) node[anchor=west] {};
        \filldraw[black] (-2,1) circle (2pt) node[anchor=west] {};
        \filldraw[black] (1,-2) circle (2pt) node[anchor=west] {};
        \filldraw[black] (-2,-2) circle (2pt) node[anchor=west] {};
        \filldraw[black] (2,2) circle (2pt) node[anchor=west] {$(\mu=2,m=2)$};
        \filldraw[black] (-2,2) circle (2pt) node[anchor=west] {};
        \filldraw[black] (2,-2) circle (2pt) node[anchor=west] {$(\mu=2,m=-2)$};
        \filldraw[black] (-1,-2) circle (2pt) node[anchor=west] {};
        \filldraw[black] (-1,2) circle (2pt) node[anchor=west] {};
        \filldraw[black] (-2,-1) circle (2pt) node[anchor=west] {};
        \filldraw[black] (2,-1) circle (2pt) node[anchor=west] {};
    \end{tikzpicture}
    \caption{\textit{Graphical View of $S^{m\mu}$ Subspaces.} Here we graphically represent each $S^{m\mu}$ subspace of RGSFs with band-limit $n_{\max}=3$ as a node. The $S^{00}$ subspace is the center red node.}\label{fig:subspace_grid}
\end{figure}
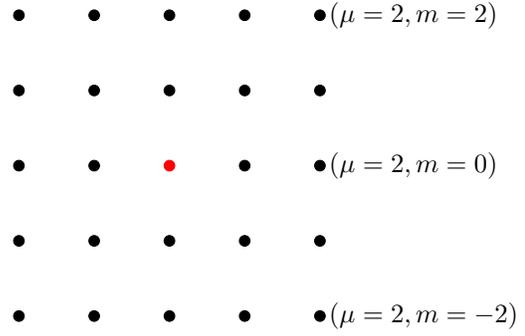

To sum the subspace dimensions it helps to visualize these subspaces first. This can be done by imagining each subspace as a point in a square grid whose side length is given by $2n_{\max}+1$. The center point is the $S^{00}$ subspace. As we move left or right $\mu$ decreases or increases, respectively. As we move up or down, $m$ increases or decreases, respectively.  In this view, see \Cref{fig:subspace_grid} where the red dot is the $S^{00}$ subspace, concentric squares around the center are all of the same dimension. So, for a given $k$, we need to sum the dimension of all of the subspaces up to the last complete square of nodes and then add the dimension of the remaining nodes in the partially used concentric square. The total dimension of all fully used squares is the same as the total number of band-limited RGSFs with $\mu,\; m \leq m_{\max}$, $N_{m_{\max}}$, where $m_{\max}$ is given by 
\begin{equation}\label{eq:side_length_formula}
    m_{\max} = \begin{cases}
        \frac{\lfloor \sqrt{k} \rfloor-1}{2}, \; \lfloor \sqrt{k} \rfloor \text{ is odd}\\
        \frac{\lfloor \sqrt{k} \rfloor - 2}{2}, \; \lfloor \sqrt{k} \rfloor \text{ is even}.
    \end{cases}
\end{equation}
To obtain \eqref{eq:side_length_formula}, we observe that each square has an odd side length, so we take the largest odd perfect square whose side length $l$ is less than or equal to $\sqrt{k}$. Then, we need to find the $m_{\max}$ that gives this side length, which is just $(l-1)/2$. Lastly, we need to sum dimensions of the remaining $S^{m\mu}$ subspaces in the $l=2(m_{\max}+1)+1$ square that is not fully used. This is $k-(2m_{\max}+1)^2$ times the dimension of the subspaces, which is $n_{\max} - m_{\max}$. Summing $(k-N_{m_{\max}})(n_{\max} - m_{\max})$ and $N_{m_{\max}}$, both with $m_{\max}$ from \eqref{eq:side_length_formula}, gives the desired result for the bound on $s$.

\subsection{Proof of Theorem \ref{thm:restricted_so3_accuracy}}\label{sec:main_res_proof}

The proof of relies on Proposition \ref{prop:rgsf_bound}. It can be seen that $\Phi_1'$ in \eqref{eq:transformed_truncated_matrix_problem} is constructed from samples of the BOS of RGSFs, $\mcl{S}'$ with the measure $\mrm{d}\rho=\mrm{d}\alpha \mrm{d}\beta \mrm{d}\gamma$ on $R$. Thus, pairing Proposition~\ref{prop:rgsf_bound} together with Theorems \ref{thm:rip_bos} and \ref{thm:sparse_recovery_rip}, with probability $1-(N_D-N_{\lambda<\lambda_c})^{-\ln^3(N_D-N_{\lambda<\lambda_c})}$, if $\widehat{a}_1$ is the solution to 
\begin{equation}
    \widehat{a}_1' = \arg \min_{z\in\mathbb{C}^{N_D}}  \|z\|_1 \; \text{subject to} \; \|Pw-P\Phi_1' z\|_2 \leq \sqrt{M}\epsilon
\end{equation}
then
\begin{equation}\label{rgsf_cs_err}
    \|a_1'-\widehat{a}_1'\|_2 \leq C_1\left( \frac{\sigma_s(a_1')_1}{\sqrt{s}} + \epsilon \right).
\end{equation}
Here $C_1>0$ only depends on the restricted isometry constant of $P\Phi_1'$, $\delta_{2s}$, and $C_2=C''C_0(\delta_{2s})^{-2}$ with $C''$ from Proposition \ref{prop:rgsf_bound} and $C_0$ from Theorem \ref{thm:rip_bos}.

Using \eqref{rgsf_cs_err} above, we can establish \eqref{eq:function_error_on_R} and \eqref{eq:D_coef_error}. First, however, we establish \eqref{eq:b_bound}.

\subsubsection{Proof of \eqref{eq:b_bound}}
The vector $b=P\Phi_2'a_2'$ has its $j$th element given by the partial series of RGSFs whose eigenvalues are less than $\lambda_c$, evaluated at $(\alpha_j,\beta_j,\gamma_j)$ and scaled by $\sqrt{\sin\beta_j}$. That is, 
\begin{equation}
    b_j = \sum_{i,\mu,m:\; \t{\lambda}{_{i}^{\mu m}} < \lambda_c} \atcoef{i}{m \mu} \sqrt{\sin \beta_j}\t{g}{_{i}^{\mu m}}(\alpha_j, \beta_j, \gamma_j),
\end{equation}
where the $\atcoef{i}{m \mu}$ are the expansion coefficients for $w(\alpha, \beta, \gamma)$ in the RGSF basis. By \eqref{eq:intermediate_rgsf_bound} and the triangle inequality we have
\begin{equation}\label{eq:mag_b}
    \left|b_j\right| \leq C'' N_D^{1/4} \sum_{i,\mu,m:\; \t{\lambda}{_{i}^{\mu m}} < \lambda_c} \left|\atcoef{i}{m \mu}\right|
\end{equation}
and the energy $E_{R^\mrm{c}}$ is 
\begin{equation}
    E_{R^\mrm{c}} = \intg{R^\mrm{c}}{}{\left|w(\alpha, \beta, \gamma)\right|^2}{\mrm{SO}(3)},
\end{equation}
or
\begin{equation}\label{eq:split_energy_integral}
    \begin{split}
        E_{R^\mrm{c}} = & \intg{\mrm{SO}(3)}{}{\left|w(\alpha, \beta, \gamma)\right|^2}{\mrm{SO}(3)} \\
        & - \intg{R}{}{\left|w(\alpha, \beta, \gamma)\right|^2}{\mrm{SO}(3)}.
    \end{split}
\end{equation}
Substituting the RGSF expansion of $w(\alpha, \beta, \gamma)$,
\begin{equation}
    w(\alpha, \beta, \gamma) = \sum_{i,\mu,m} \atcoef{i}{m \mu}\t{g}{_{i}^{\mu m}}(\alpha_j, \beta_j, \gamma_j),
\end{equation}
into \eqref{eq:split_energy_integral} and using \eqref{eq:SO3_orthogonality} and \eqref{eq:R_orthogonality} yields 
\begin{equation}
    E_{R^\mrm{c}} = \sum_{i,\mu,m} \left(1-\t{\lambda}{_{i}^{\mu m}}\right)\left| \atcoef{i}{m \mu} \right|^2.
\end{equation}
If we restrict the sum above to only those coefficients on RGSFs with eigenvalues less than $\lambda_c$ we have 
\begin{equation}\label{eq:energy_bound}
    E_{R^\mrm{c}} \geq \sum_{i,\mu,m:\; \t{\lambda}{_{i}^{\mu m}} < \lambda_c} \left(1-\t{\lambda}{_{i}^{\mu m}}\right) \left| \atcoef{i}{m \mu} \right|^2
\end{equation}
and using the fact that $1-\lambda_c<1-\t{\lambda}{_{i}^{\mu m}}$ and dividing by $1-\lambda_c$ gives 
\begin{equation}\label{eq:2norm_of_coefs_less_lam}
    \frac{E_{R^\mrm{c}}}{(1-\lambda_c)} \geq \sum_{i,\mu,m:\; \t{\lambda}{_{i}^{\mu m}} < \lambda_c}  \left| \atcoef{i}{m \mu} \right|^2.
\end{equation}
Noting the sum in \eqref{eq:mag_b} is the $\ell_1$ norm of the vector of coefficients on the RGSFs with eigenvalues less than $\lambda_c$ and that \eqref{eq:2norm_of_coefs_less_lam} is the $\ell_2$ norm of the same vector. Then, given the fact that for $v\in \mathbb{C}^\mcl{N}$, $\|v\|_1 \leq \sqrt{\mcl{N}}\|v\|_2$, we find
\begin{equation}
    \left|b_j\right| \leq \frac{C'' N_D^{1/4} \sqrt{N_{\lambda<\lambda_c}} E_{R^\mrm{c}}}{(1-\lambda_c)},
\end{equation}
where $N_{\lambda<\lambda_c}$ is the number of RGSF eigenvalues less than $\lambda_c$. Thus, since $N_{\lambda<\lambda_c}\leq N$, we have
\begin{equation}
    \|b\|_\infty \leq \frac{C'' N_D^{3/4} E_{R^\mrm{c}}}{(1-\lambda_c)}.
\end{equation}

\subsubsection{Proof of \eqref{eq:function_error_on_R}}
We begin with the explicit calculation of the error,
\begin{equation}
    \begin{split}
        \|w(\alpha, \beta, \gamma) - & \widehat{w}(\alpha, \beta, \gamma)\|_R^2 = \\
        & \sum_{i,\mu,m}\sum_{i',\mu',m'}  \t{\chi}{_i^{m\mu}} \overline{\t{\chi}{_{i'}^{m'\mu'}}}  \inner{\t{g}{_{i}^{\mu m}}}{\t{g}{_{i'}^{\mu' m'}}}{R},
    \end{split}
\end{equation}
where $\t{\chi}{_i^{m\mu}}=\atcoef{i}{m \mu}- \ahtcoef{i}{m \mu}$. By the orthogonality of the RGSFs on $R$ we find
\begin{equation}\label{eq:w_error_as_sum}
    \|w(\alpha, \beta, \gamma) - &\widehat{w}(\alpha, \beta, \gamma)\|_R^2 = \sum_{i,\mu,m}\t{\lambda}{_{i}^{\mu m}} \left|\t{\chi}{_i^{m\mu}}\right|^2.
\end{equation}
The right side of \eqref{eq:w_error_as_sum} can be rewritten in terms of $a_1'$ and $\widehat{a}_1'$ by splitting the sum into two parts and absorbing the RGSF eigenvalue into the absolute values:
\begin{equation}
    \begin{split}
        \sum_{i,\mu,m}\t{\lambda}{_{i}^{\mu m}} \left|\t{\chi}{_i^{m\mu}}\right|^2 & = \sum_{i,\mu,m}\t{\lambda}{_{i}^{\mu m}} \left|\atcoef{i}{m \mu}- \ahtcoef{i}{m \mu}\right|^2 \\
        =& \sum_{i,\mu,m:\t{\lambda}{_{i}^{\mu m}}\geq\lambda_c} \left|\sqrt{\t{\lambda}{_{i}^{\mu m}}}\atcoef{i}{m \mu} - \sqrt{\t{\lambda}{_{i}^{\mu m}}}\ahtcoef{i}{m \mu}\right|^2 \\ 
        & \quad + \sum_{i,\mu,m:\t{\lambda}{_{i}^{\mu m}}<\lambda_c} \t{\lambda}{_{i}^{\mu m}}\left|\atcoef{i}{m \mu} - \ahtcoef{i}{m \mu}\right|^2 \\
        = &  \|a_1'-\widehat{a}_1'\|^2 \\
        &+ \sum_{i,\mu,m:\t{\lambda}{_{i}^{\mu m}}<\lambda_c} \t{\lambda}{_{i}^{\mu m}}\left|\atcoef{i}{m \mu} - \ahtcoef{i}{m \mu}\right|^2.
    \end{split}
\end{equation}
The coefficients in $\widehat{a}'_2$ are zero, so the $\ahtcoef{i}{m \mu}$ on the right can be set to zero. Thus,
\begin{equation}
    \begin{split}
        \|w(\alpha, \beta, \gamma) - & \widehat{w}(\alpha, \beta, \gamma)\|_R^2= \\
        & \|a_1'-\widehat{a}_1'\|^2 + \sum_{i,\mu,m:\t{\lambda}{_{i}^{\mu m}}<\lambda_c} \t{\lambda}{_{i}^{\mu m}}\left|\atcoef{i}{m \mu} \right|^2.
    \end{split}
\end{equation} 
Since $\t{\lambda}{_{i}^{\mu m}}<\lambda_c$ in the remaining sum, we have the bound 
\begin{equation}
    \begin{split}
        \|w(\alpha, \beta, \gamma) - &\widehat{w}(\alpha, \beta, \gamma)\|_R^2 < \\
        & \|a_1'-\widehat{a}_1'\|^2 + \lambda_c \sum_{i,\mu,m:\t{\lambda}{_{i}^{\mu m}}<\lambda_c} \left|\atcoef{i}{m \mu} \right|^2.
    \end{split}
\end{equation}
We can then use \eqref{eq:energy_bound} and the error bound in \eqref{rgsf_cs_err} to find the desired result, namely,
\begin{equation}
    \begin{split}
        \|w(\alpha, \beta, \gamma) - & \widehat{w}(\alpha, \beta, \gamma)\|_R^2 < \\
        &C_1^2\left( \frac{\sigma_s(a_1')_1}{\sqrt{s}} + \epsilon \right)^2 + \frac{\lambda_c}{1-\lambda_c} E_{R^\mrm{c}}
    \end{split}
\end{equation}
with probability as stated in Theorem \ref{thm:restricted_so3_accuracy}.

\subsubsection{Proof of \eqref{eq:D_coef_error}}
We begin by noting the Parseval relation for a function $w(\alpha, \beta, \gamma)$ that can be written as a series of band-limited Wigner $D$-functions with coefficients $a$, that is $\left\|w(\alpha, \beta, \gamma)\right\|_{\sot}^2=\|a\|^2$. This implies for $w(\alpha, \beta, \gamma)$ and $\widehat{w}(\alpha, \beta, \gamma)$ we have 
\begin{equation}
    \|a-\widehat{a}\|^2 = \left\|w(\alpha, \beta, \gamma)-\widehat{w}(\alpha, \beta, \gamma)\right\|_{\sot}^2.
\end{equation}
We can rewrite the right hand side by splitting the norm into a sum of $\|\cdot\|_R^2$ and $\|\cdot\|_{R^\mrm{c}}^2$ and this gives
\begin{equation}
    \begin{split}
        \|a-\widehat{a}\|^2 = & \left\|w(\alpha, \beta, \gamma)-\widehat{w}(\alpha, \beta, \gamma)\right\|_R^2  \\ 
        &+ \left\|w(\alpha, \beta, \gamma)-\widehat{w}(\alpha, \beta, \gamma)\right\|_{R^\mrm{c}}^2.
    \end{split}
\end{equation}
By the triangle inequality we obtain
\begin{equation}
    \begin{split}
        \|a-\widehat{a}\|^2 \leq & \left\|w(\alpha, \beta, \gamma)-\widehat{w}(\alpha, \beta, \gamma)\right\|_R^2 \\
        & + \left(\left\|w(\alpha, \beta, \gamma)\right\|_{R^\mrm{c}} + \left\|\widehat{w}(\alpha, \beta, \gamma)\right\|_{R^\mrm{c}}\right)^2.
    \end{split}
\end{equation}
Noting the two terms in the parentheses are $\sqrt{E_{R^\mrm{c}}}$ and $\sqrt{\widehat{E}_{R^\mrm{c}}}$ and using \eqref{eq:function_error_on_R} gives the desired result.

\section{SW Coefficients Used in Numerical Examples}
The SW coefficients used for the numerical examples in the main manuscript are given below in \Cref{tbl:sw_coef}.
\begin{table}[h]
\centering
    \begin{tabular}{c|c}
        $n$ & $a_n^0$ \\
        \hline
        0 & $0.0092 + 0.0924\mi$ \\
        1 & $-0.1249 - 0.0090\mi$ \\
        2 & $0.0047 - 0.1245\mi$ \\
        3 & $0.1271 - 0.0126\mi$ \\
        4 & $0.0155 + 0.1290\mi$ \\
        5 & $-0.1273 + 0.0173\mi$ \\
        6 & $-0.0280 - 0.1241\mi$ \\
        7 & $0.1169 - 0.0419\mi$ \\
        8 & $0.0542 + 0.1068\mi$ \\
        9 & $-0.0954 + 0.0624\mi$ \\
        10 & $-0.0680 - 0.0790\mi$ \\
        11 & $0.0590 - 0.0751\mi$ \\
        12 & $0.0799 + 0.0400\mi$ \\
        13 & $-0.0189 + 0.0745\mi$ \\
        14 & $-0.0633 + 0.0032\mi$ \\
        15 & $-0.0169 - 0.0507\mi$ \\
        16 & $0.0323 - 0.0218\mi$ \\
        17 & $0.0225 + 0.0167\mi$ \\
        18 & $-0.0073 + 0.0195\mi$ \\
        19 & $-0.0136 - 0.0010\mi$ \\
        20 & $-0.0016 - 0.0087\mi$
    \end{tabular}
    \caption{Field Coefficients for the acoustic field used in the numerical examples in Compressive Sensing with Wigner $D$-functions on Subsets of the Sphere.}\label{tbl:sw_coef}
\end{table}

%
%
%
%

%


%


\ifCLASSOPTIONcaptionsoff
  \newpage
\fi